\documentclass[prd,nofootinbib,english,notitlepage,a4paper]{revtex4-1}
\usepackage{amsmath}
\usepackage{amsfonts,color}
\usepackage{amssymb,float}
\usepackage[utf8]{inputenc}
\usepackage[T1]{fontenc}
\usepackage[svgnames]{xcolor}
\usepackage[unicode=true,pdfusetitle,
 bookmarks=true,bookmarksnumbered=true,bookmarksopen=true,bookmarksopenlevel=3,
 breaklinks=false,pdfborder={0 0 1},backref=false,colorlinks=true]
 {hyperref}
\usepackage{enumitem}

\newcommand{\udt}[3]{#1^{#2}_{\phantom{#2}#3}}
\newcommand{\udut}[4]{#1^{#2\phantom{#3}#4}_{\phantom{#2}#3\phantom{#4}}}

\newcommand{\dut}[3]{#1_{#2}^{\phantom{#2}#3}}
\newcommand{\dudt}[4]{#1_{#2\phantom{#3}#4}^{\phantom{#2}#3}}

\newcommand{\order}[2]{\overset{\scriptscriptstyle #2}{#1}\vphantom{#1}}

\newcommand{\GG}{\mathbf{G}}
\newcommand{\FF}{\mathbf{F}}
\newcommand{\HH}{\mathbf{H}}
\newcommand{\LL}{\mathbf{L}}

\begin{document}
\title{Post-Newtonian limit of Teleparallel Horndeski gravity}

\author{Sebastian Bahamonde}
\email{sbahamonde@ut.ee, sebastian.beltran.14@ucl.ac.uk}
\affiliation{Laboratory of Theoretical Physics, Institute of Physics, University of Tartu, W. Ostwaldi 1, 50411 Tartu, Estonia}
\affiliation{Department of Mathematics, University College London, Gower Street, London, WC1E 6BT, United Kingdom}

\author{Konstantinos F.	Dialektopoulos}
\email{kdialekt@gmail.com}
\affiliation{Center for Gravitation and Cosmology, College of Physical Science and Technology, Yangzhou University, Yangzhou 225009, China}

\author{Manuel Hohmann}
\email[Corresponding Author: ]{manuel.hohmann@ut.ee}
\affiliation{Laboratory of Theoretical Physics, Institute of Physics, University of Tartu, W. Ostwaldi 1, 50411 Tartu, Estonia}

\author{Jackson Levi Said}
\email{jackson.said@um.edu.mt}
\affiliation{Institute of Space Sciences and Astronomy, University of Malta, Msida, MSD 2080, Malta}
\affiliation{Department of Physics, University of Malta, Msida, MSD 2080, Malta}

\begin{abstract}
We consider the newly proposed Bahamonde-Dialektopoulos-Levi Said (BDLS) theory, that is the Horndeski analog in the teleparallel framework and thus contains a non-minimally coupled scalar field, including higher order derivatives, that leads however to second order field equations both for the tetrad and the scalar field. This theory was mostly constructed to revive those models that were severely constrained in the scalar-tensor version of the theory from the GW170817, but includes also much richer phenomenology because of the nature of the torsion tensor. For this theory we determine the parametrized post-Newtonian limit, calculate the full set of post-Newtonian parameters and highlight some special cases.
\end{abstract}

\maketitle

\section{Introduction}

General Relativity (GR) and its cosmological model, $\Lambda$CDM, are known to possess some features that do not go along with observations \cite{Aghanim:2018eyx,Riess:2018kzi}. The accelerating expansion of the Universe is usually associated to the cosmological constant $\Lambda$, the observed value of which differs from the theoretical prediction, i.e. considered as the vacuum energy from Quantum Field Theory (QFT),  for 120 orders of magnitude. On top of this, experiments fail to detect a suitable particle candidate for dark matter (an unknown form of matter that interacts only gravitationally), as well as a TeV-scale supersymmetry. These problems, together with some other astrophysical ones, motivated physicists to start formulating alternative descriptions of the gravitational interactions, leading to a plethora of modified theories of gravity \cite{Clifton:2011jh,Capozziello:2011et,Nojiri:2017ncd}.

Many theories in several different contexts have been studied throughout the years in the literature. Some indicative examples are extensions of GR such as $f(\mathring{R})$, $f(\mathcal{\mathring{G}})$ models \cite{Sotiriou:2008rp,Nojiri:2010wj,Bahamonde:2019swy,Bajardi:2019zzs}; theories involving extra fields like scalar-tensor theories, Tensor-Vector-Scalar (TeVeS) theories, Einstein-Aether and so on; higher dimensional theories like the Dvali-Gabadadze-Porratti (DGP) model, Kaluza-Klein, Randall-Sundrum I \& II and more \cite{Clifton:2011jh}; as well as non-local theories involving terms as $\mathring{\square} ^{-1} \mathring{R}$ and other such scalars\footnote{The overcircle $\mathring{}$ refers to quantities calculated with the Levi-Civita connection. More details about the notation will be given later on.} \cite{Deser:2007jk,Modesto:2014lga,Bahamonde:2017bps,Bahamonde:2017sdo}.

Lately, it has been realised that gravitational interactions can be equivalently described by three different theories \cite{BeltranJimenez:2019tjy}. This is known in the literature as the \textit{Geometric Trinity of Gravity} and it refers to three theories all describing gravity in a different mathematical way, using different connections of the spacetime, but are all equivalent with each other \cite{Koivisto:2018loq,BeltranJimenez:2018vdo}. The first one is GR that is based on the curvature of the Levi-Civita connection, the second one is Teleparallel gravity (TG) \cite{Aldrovandi:2013wha,Krssak:2018ywd} that is based on the torsion of the Weitzenb\"ock connection, and the third one is Symmetric Teleparallel gravity (STG) \cite{Jimenez:2019yyx,Conroy:2017yln,BeltranJimenez:2017tkd} based on the non-metricity of a flat, symmetric connection. Even in these alternative formulations of GR based flat connections, there have been proposed many modifications \cite{Krssak:2018ywd,Cai:2015emx,Jimenez:2019ovq,Dialektopoulos:2019mtr,Hohmann:2018vle,Hohmann:2018dqh,Bahamonde:2018miw}.

Horndeski theory is the most general scalar-tensor theory with a single scalar field in four dimensions leading to second order field equations \cite{1974IJTP...10..363H}. Recently, part of the authors proposed a new theory, that is a reformulation of Horndeski theory in the Teleparallel framework, i.e. using a connection that is curvature- and nonmetricity-free and possesses only torsion \cite{Bahamonde:2019shr}. Based on that, S.~Bahamonde, K.~F.~Dialektopoulos and J.~Levi Said (from hereon BDLS) introduced a scalar field and constructed the Horndeski analog in this ``different'' geometry. The resulting theory should have second order field equations, in order to avoid ghosts, should not contain parity-violating terms and up to quadratic contractions of the torsion tensor. The motivation for this, is the fact that Horndeski theory in its known, i.e. curvature, formulation, was severely constrained by the GW170817 event. It turns out that, in the teleparallel framework not only the eliminated couplings revive,  but also many more interesting ones result because of the presence of a new function \cite{Bahamonde:2019ipm}.

There are several ways to constrain modifications of gravity. Theoretically, they should satisfy some  criteria such as positivity of energy, causal structure and so on. Another important criterion is to confront the theory with observations. To do so, we have to bring the theory to a form that is characterized by a certain amount of parameters and to compare these parameters with high precision measurements of their values in the solar system in order to constrain different classes of theories. A broadly used and well established tool to test modifications of gravity is the parametrized Post-Newtonian (PPN) formalism which effectively characterizes gravity theories by a set of ten parameters and by comparing them with high precision data from the Solar System, we can study the viability of the theory.

Several studies of the PPN formalism in scalar-torsion theories have shown that many models give the same values with GR to the PPN parameters and thus cannot be distinguished using these high-precision measurements. More specifically, in \cite{Li:2013oef} and \cite{Chen:2014qsa} the authors study the PPN expansion of teleparallel dark energy models showing that they cannot be distinguished from GR and thus being much different than scalar-tensor theories \cite{Hohmann:2013rba,Scharer:2014kya} that need specific screening mechanisms to pass Solar system tests. In \cite{Sadjadi:2016kwj} they study the same models as before by adding a non-minimal coupling of the scalar field with the boundary term. Interestingly enough, they find that this new coupling affects the PPN parameters of the theory, which they obtain for different cases. The interested reader should also check the recent works on the subject \cite{Flathmann:2019khc,Emtsova:2019qsl,Ualikhanova:2019ygl}. In this paper, we study the PPN parametrization of the BDLS theory, i.e. the teleparallel analog of Horndeski gravity, in order to constrain its parameters from various experiments.

This paper is organized as follows: in section \ref{sec:Teleparallel} we review the basics of Teleparallel gravity, meaning the fundamental fields of the theory and the basic underlying principles; in addition, we formulate the recently proposed BDLS theory. In Sec.~\ref{sec:HorndeskiTeleparallel} we derive the field equations for this theory both for the tetrad and for the scalar field. The main result of the paper is presented in Sec.~\ref{sec:PPN} where we discuss the PPN expansion of BDLS theory. Finally, in Sec.~\ref{sec:special} we discuss some special cases, i.e. models that draw more attention in the literature. In Sec.~\ref{sec:Conclusions} we conclude our results and discuss future aspects.

Throughout the paper, capital Latin letters $A,B,C,...$ are Lorentz indices, while the Greek ones $\alpha,\beta,\mu,...$ represent coordinates of the spacetime manifold. Furthermore, small Latin letters from the middle of the alphabet, i.e. $i, j, k, ...$ are used for spatial indices. Quantities calculated with the Levi-Civita connection (e.g. connections, covariant derivatives, d'Alembertians) are given with a circle on top, e.g. $\mathring{\nabla}_{\mu}$ and quantities referring to flat spacetime are denoted with a bar on top, e.g. $\bar{\Box}$. All the other quantities that have no symbols, e.g. $\Gamma ^{\alpha}{}_{\mu\nu}$, are calculated with (or referred to) the Weitzenb\"ock connection. Also unless otherwise stated, we use the metric signature $\eta_{\mu\nu}=\textrm{diag}(-1,1,1,1)$, and geometric units.

\section{Teleparallel gravity and its Extension to BDLS Theory}\label{sec:Teleparallel}
Through the Levi-Civita connection, $\mathring{\Gamma}_{\mu\nu}^{\sigma}$, GR expresses gravitation as geometric curvature. In curvature-based theories of gravity, the amount of curvature present in a system is then expressed through the Riemann tensor which is the fundamental measure of curvature in standard gravity \cite{misner1973gravitation}. This has been extended to produce several extended theories of gravity \cite{Clifton:2011jh,Capozziello:2011et} which are constructed on the Levi-Civita connection together with the metric tensor.

On the other hand, the fundamental dynamical object of TG is the tetrad, $\udt{e}{A}{\mu}$, which acts as a soldering agent \cite{Aldrovandi:2013wha} between the general manifold (Greek indices) and Minkowski space (capital Latin indices). Along this line of reasoning, the tetrad can readily be used to raise Minkowski space indices to the general manifold or vice versa \cite{RevModPhys.48.393}
\begin{align}
    g_{\mu\nu} &= \udt{e}{A}{\mu}\udt{e}{B}{\nu} \eta_{AB}\,,\\
    \eta_{AB} &= \dut{E}{A}{\mu}\dut{E}{B}{\nu} g_{\mu\nu}\,,
\end{align}
which serve as the definition for the inverse tetrad, $\dut{E}{A}{\mu}$, which must also adhere to the orthonormality conditions
\begin{align}
    \udt{e}{A}{\mu}\dut{E}{B}{\mu} &= \delta_B^A\,,\\
    \udt{e}{A}{\mu}\dut{E}{A}{\nu} &= \delta_{\mu}^{\nu}\,,
\end{align}
which also normalise the tetrads. However, there are an infinite number of tetrad choices that satisfy these conditions due to the local Lorentz transformations (LLTs), $\udt{\Lambda}{A}{B}$, on the Minkowski  space.

TG is based on the replacement of the Levi-Civita connection with a flat, metric connection which exchanges the tetrad as the fundamental dynamical variable instead of the metric. This is the most general linear affine connection that is curvatureless and satisfies metricity $\nabla_\mu g_{\alpha\beta}=0$~\cite{Weitzenbock1923}. This connection can be expressed as \cite{Cai:2015emx,Krssak:2018ywd}
\begin{equation}
\Gamma^{\sigma}{}_{\nu\mu} = \dut{E}{A}{\sigma}\partial_{\mu}\udt{e}{A}{\nu} + \dut{E}{A}{\sigma}\udt{\omega}{A}{B\mu}\udt{e}{B}{\nu}\,,
\end{equation}
where $\udt{\omega}{A}{B\mu}$ represents the components of a flat spin connection, \(\partial_{[\mu}\omega^A{}_{|B|\nu]} + \omega^A{}_{C[\mu}\omega^C{}_{|B|\nu]} \equiv 0\). As in GR, this object accounts for the LLT degrees of freedom (DoF), but unlike GR, this is a flat connection and plays an active role in the field equations to counter any inertial effects from the LLT invariance of the theory. Naturally, there always will exist a frame in which the spin connection components will be allowed to vanish, this is the so-called purely inertial or Weitzenböck gauge \cite{Krssak:2018ywd}.

In choosing the Weitzenb\"{o}ck connection, the Riemann tensor turns out to identically vanish for every choice of tetrad (or the metric it produces) irrespective of its components. This occurs because the Riemann tensor measures curvature which is associated with the connection not the metric and so TG necessitates another way to measure geometric deformation due to gravitation. To do this, TG uses the torsion tensor defined by the antisymmetry of the connection through \cite{Aldrovandi:2013wha,ortin2004gravity}
\begin{equation}
\udt{T}{A}{\mu\nu} := 2\Gamma^A{}_{[\nu\mu]}\,,
\end{equation}
which is a measure of the field strength of gravitation in TG, and where square brackets denote antisymmetry. This quantity transforms covariantly under local Lorentz transformations~\cite{Krssak:2015oua}. The torsion tensor can be readily decomposed into irreducible axial, vector and purely tensorial parts which are defined by \cite{PhysRevD.19.3524,Bahamonde:2017wwk}
\begin{align}
    a_{\mu} &:= \frac{1}{6}\epsilon_{\mu\nu\sigma\rho}T^{\nu\sigma\rho}\,,\\
    v_{\mu} &:= \udt{T}{\sigma}{\sigma\mu}\,,\\
    t_{\sigma\mu\nu} &:= \frac{1}{2}\left(T_{\sigma\mu\nu} + T_{\mu\sigma\nu}\right) + \frac{1}{6}\left(g_{\nu\sigma} v_{\mu} + g_{\nu\mu}v_{\sigma}\right) - \frac{1}{3}g_{\sigma\mu}v_{\nu}\,,
\end{align}
where $\epsilon_{\mu\nu\sigma\rho}$ represents the totally antisymmetric Levi-Civita symbols in four dimensions, and where here and in the remainder of this article we use the tetrad and its inverse to implicitly translate between Lorentz and spacetime indices, \(T^{\rho}{}_{\mu\nu} = E_A{}^{\rho}T^A{}_{\mu\nu}\). These tensors are irreducible parts with respect to the local Lorentz group, and vanish when contracted with each other due to the symmetries of the torsion tensor. The axial, vector, and purely tensorial parts can be used to construct the scalar invariants
\begin{align}
    T_{\text{ax}} &:= a_{\mu}a^{\mu} = \frac{1}{18}\left(T_{\sigma\mu\nu}T^{\sigma\mu\nu} - 2T_{\sigma\mu\nu}T^{\mu\sigma\nu}\right)\,,\\
    T_{\text{vec}} &:= v_{\mu}v^{\mu} = \udt{T}{\sigma}{\sigma\mu}\dut{T}{\rho}{\rho\mu}\,,\\
    T{_\text{ten}} &:= t_{\sigma\mu\nu}t^{\sigma\mu\nu} = \frac{1}{2}\left(T_{\sigma\mu\nu}T^{\sigma\mu\nu} + T_{\sigma\mu\nu}T^{\mu\sigma\nu}\right) - \frac{1}{2}\udt{T}{\sigma}{\sigma\mu}\dut{T}{\rho}{\rho\mu}\,,
\end{align}
which are all possible scalar invariants that are parity preserving that can be produced from these irreducible parts \cite{Bahamonde:2015zma}. In fact, these scalar invariants form the most general purely gravitational Lagrangian, $f(T_{\text{ax}},T_{\text{vec}},T{_\text{ten}})$~\cite{Bahamonde:2017wwk}, that is quadratic in torsion in that the scalars are at most quadratic and second-order in terms of the resulting field equations, while not being parity violating. Another critical feature of these scalars is that the linear combination, labelled as the torsion scalar,
\begin{equation}
    T := \frac{3}{2} T_{\text{ax}} + \frac{2}{3} T_{\text{ten}} - \frac{2}{3} T{_\text{vec}}=\frac{1}{2} \left(E_A{}^\sigma g^{\rho \mu} E_B{}^\nu + 2 E_B{}^\rho g^{\sigma \mu} E_A{}^\nu + \frac{1}{2} \eta_{AB} g^{\mu\rho} g ^{\nu\sigma} \right) T^A{}_{\mu\nu} T^B{}_{\rho\sigma}\,,
\end{equation}
turns out to be equal to the regular Ricci scalar, $\mathring{R}$ (calculated using the Levi-Civita connection), up to a total divergence term given by \cite{Bahamonde:2015zma}
\begin{equation}
    R=\mathring{R}+T-\frac{2}{e}\partial_{\mu}\left(e \udut{T}{\sigma}{\sigma}{\mu}\right) = 0\,,
\end{equation}
where $R$ is the Ricci scalar determined with the Weitzenb\"{o}ck connection which naturally vanishes, and $e=\text{det}\left(\udt{e}{A}{\mu}\right)=\sqrt{-g}$ is the tetrad determinant. This means that
\begin{equation}
    \mathring{R} = -T+\frac{2}{e}\partial_{\mu}\left(e \udut{T}{\sigma}{\sigma}{\mu}\right) :=-T+B\,.
\end{equation}
Given that $B$ is a boundary term, a torsion scalar Lagrangian will produce identical field equations as those of GR and thus form the Teleparallel Gravity equivalent to General Relativity (TEGR) \cite{Hehl:1994ue,Aldrovandi:2013wha}, despite the total divergence difference at the level of the Lagrangian.

In GR, the procedure by which local Lorentz frames are transformed to general ones is by the exchange of the Minkowski metric with its general metric tensor while also raising the partial derivative to the covariant derivative associated with the Levi-Civita connection \cite{misner1973gravitation}. TG is different in that this coupling prescription is guided by an exchange of so-called trivial (tangent space) tetrads with their general manifold tetrad analog, while, for a scalar field $\Psi=\Psi(x)$, the identical derivative procedure is kept, i.e. \cite{Aldrovandi:2013wha}
\begin{equation}
    \partial_{\mu} \rightarrow \mathring{\nabla}_{\mu}\,,
\end{equation}
which emphasises the close relationship both theories have \cite{BeltranJimenez:2019tjy}.

With both the gravitational and scalar field sections adequately developed, we consider the conditions on which the Teleparallel Gravity analog of the Horndeski framework in four dimensions is built \cite{Bahamonde:2019shr}, which are (i) the dynamical equations of the theory are at most second-order in their derivatives of the tetrads; (ii) the scalar invariants are not parity violating; and (iii) at most quadratic contractions of the torsion tensor are allowed. In standard gravity, the Lovelock theorem \cite{Lovelock:1971yv} shows that any Lagrangian beyond that in the Einstein-Hilbert action (up to a constant) cannot remain second-order in their field equations. This is not the case in TG \cite{Gonzalez:2015sha} where a potentially infinite number of terms can be incorporated into the Lagrangian of a second-order theory. Condition (iii) is a statement about the possible terms that are considered from this infinite series of terms, where higher order corrections may play a role in other phenomenology.

These conditions directly produce a finite set of scalar invariants that describe the nonminimal coupling with the scalar field for the linear appearance of the torsion tensor \cite{Bahamonde:2019shr}
\begin{equation}
    I_2 = v^{\mu} \phi_{;\mu}\,,
\end{equation}
and the quadratic torsion tensor coupling terms
\begin{align}
    J_1 &= a^{\mu}a^{\nu} \phi_{;\mu}\phi_{;\nu}\,,\\
    J_3 &= v_{\sigma} t^{\sigma\mu\nu}\phi_{;\mu}\phi_{;\nu}\,,\\
    J_5 &= t^{\sigma\mu\nu}\dudt{t}{\sigma}{\bar{\mu}}{\nu}\phi_{;\mu}\phi_{;\bar{\mu}}\,,\\
    J_6 &= t^{\sigma\mu\nu}\dut{t}{\sigma}{\bar{\mu}\bar{\nu}}\phi_{;\mu}\phi_{;\nu}\phi_{;\bar{\mu}}\phi_{;\bar{\nu}}\,,\\
    J_8 &= t^{\sigma\mu\nu}\dut{t}{\sigma\mu}{\bar{\nu}}\phi_{;\nu}\phi_{;\bar{\nu}}\,,\\
    J_{10} &= \udt{\epsilon}{\mu}{\nu\sigma\rho}a^{\nu}t^{\alpha\rho\sigma}\phi_{;\mu}\phi_{;\alpha}\,,
\end{align}
which also observe the other conditions, and where semicolon represents Levi-Civita covariant derivatives.

These new scalar invariants can be arbitrarily combined to produce a new Lagrangian term
\begin{equation}
    \mathcal{L}_{\text{Tele}}:= G_{\text{Tele}}\left(\phi,X,T,T_{\text{ax}},T_{\text{vec}},I_2,J_1,J_3,J_5,J_6,J_8,J_{10}\right)\,,
\end{equation}
where the kinetic term is defined as $X:=-\frac{1}{2}\partial^{\mu}\phi\partial_{\mu}\phi$. Given the lower-order nature of TG, the other terms of standard Horndeski gravity retain their original formulations except that they are now expressed through the tetrad formalism. Therefore the TG analog of Horndeski' theory of gravity \cite{Horndeski:1974wa} turns out to be \cite{Bahamonde:2019shr}
\begin{equation}\label{action}
    \mathcal{S}_{\text{BDLS}} = \frac{1}{2\kappa^2}\int d^4 x\, e\mathcal{L}_{\text{Tele}} + \frac{1}{2\kappa^2}\sum_{i=2}^{5} \int d^4 x\, e\mathcal{L}_i+ \int d^4x \, e\mathcal{L}_{\rm m}\,,
\end{equation}
where
\begin{align}
    \mathcal{L}_2 &:= G_2(\phi,X)\,,\\
    \mathcal{L}_3 &:= G_3(\phi,X)\mathring{\Box}\phi\,,\\
    \mathcal{L}_4 &:= G_4(\phi,X) \left(-T+B\right) + G_{4,X}(\phi,X)\left[\left(\mathring{\Box}\phi\right)^2 - \phi_{;\mu\nu}\phi^{;\mu\nu}\right]\,,\\
    \mathcal{L}_{5} &:= G_5(\phi,X)\mathring{G}_{\mu\nu}\phi^{;\mu\nu} - \frac{1}{6}G_{5,X}(\phi,X)\left[\left(\mathring{\Box}\phi\right)^3 + 2\dut{\phi}{;\mu}{\nu}\dut{\phi}{;\nu}{\alpha}\dut{\phi}{;\alpha}{\mu} - 3\phi_{;\mu\nu}\phi^{;\mu\nu}\,\mathring{\Box}\phi\right]\,,
\end{align}
with $\mathcal{L}_{\rm m}$ being the matter Lagrangian in the Jordan conformal frame, $\kappa^2=8\pi G$, $\mathring{G}_{\mu\nu}$ is the standard Einstein tensor, and comma represents partial derivatives. The standard Horndeski gravity is clearly recovered for the choice $G_{\text{Tele}} = 0$. Stemming from the invariance under LLTs and the general covariance of the underlying torsion tensor formalism, the TG analog of Horndeski's theory of gravity turns out to also adhere to these invariance properties. Let us emphasise here that since the torsion tensor is covariant under local Lorentz transformations, the BDLS theory is also invariant under these transformations~\cite{Bahamonde:2019shr}.

\section{Field equations for BDLS theory}
\label{sec:HorndeskiTeleparallel}
In this section, we will present the field equations in BDLS gravity. By varying the action~\eqref{action} with respect to the tetrads, we find  that
\begin{eqnarray}
 \delta_{e}\mathcal{S}_{\rm BDLS}= e\mathcal{L}_{\rm Tele}E_A{}^{\mu}\delta e^A{}_\mu+e\delta_{e} \mathcal{L}_{\rm Tele}+e\sum_{i=2}^{5}\mathcal{L}_iE_A{}^{\mu}\delta e^A{}_\mu+e \delta_{e} \sum_{i=2}^{5}\mathcal{L}_{i}+2\kappa^2e\Theta_A{}^\mu\delta e^{A}{}_\mu=0\,,\label{deltaS}
\end{eqnarray}
where we have used $\delta e=e E_A{}^{\mu} \delta e^A{}_\mu$, and we have defined the energy-momentum tensor as
\begin{align}
\Theta_A{}^\mu=\frac{1}{e}\frac{\delta (e\mathcal{L}_{\rm m})}{\delta e^A{}_{\mu}}\,.
\end{align}
We assume that the matter action is Lorentz invariant, and that all matter fields are minimally coupled to the metric only, from which follows that the energy-momentum tensor \(\Theta_{\mu\nu} = e^A{}_{\mu}g_{\nu\rho}\Theta_A{}^{\rho}\) is symmetric, \(\Theta_{[\mu\nu]} = 0\). Hence, all test matter follows the geodesics of the metric tensor, thus satisfying the weak equivalence principle. It should be noted that throughout this paper we use the standard notation for antisymmetrization of tensors. For example, for a rank-2 tensor $A_{[ij]}=(A_{ij}-A_{ji})/2$.

The variations of $\delta_{e}\sum_{i=2}^{5} \mathcal{L}_i$ gives the standard Horndeski field equations whereas the variations $\delta_{e} \mathcal{L}_{\rm Tele}$ are related to the extra terms coming from Teleparallel gravity. After doing several computations, one finds that the field equations can be written as
\begin{eqnarray}
&&4(\partial_{\lambda}G_{\rm Tele,T})S_{A}\,^{\lambda\mu}+4e^{-1}\partial_{\lambda}(e S_{A}\,^{\lambda\mu})G_{\rm Tele,T}-4G_{\textrm{Tele},T}T^{\sigma}\,_{\lambda A}S_{\sigma}\,^{\mu\lambda}+4G_{\rm Tele,T}\omega^{B}{}_{A\nu}S_{B}{}^{\nu\mu}\nonumber\\
&&-\phi_{;A}\Big[G_{\rm Tele,X}\phi^{;\mu}-G_{\rm Tele,I_2}v^\mu   -2G_{\rm Tele,J_1}a^{\mu}a_{J}\phi^{;J} +G_{\rm Tele,J_3}v_I t_{K}{}^{\mu I}\phi^{;K}-2G_{\rm Tele,J_5} t^{I\mu K}t_{IJK}\phi^{;J}\nonumber\\
&&+2G_{\rm Tele,J_6}t_{ILK}t^\mu{}_M{}^I\phi^{;K}\phi^{;L}\phi^{;M}-2G_{\rm Tele,J_8}t_{IJK}t^{IJ}{}^{\mu}\phi^{;K}-G_{\rm Tele,J_{10}}a^J \phi^{;I}\Big(\epsilon_{\mu JCD}t_I{}^{CD}+\epsilon_{IJCD}t^{\mu CD}\Big)  \Big]\nonumber\\
&&+\frac{1}{3}\Big[M^I(\epsilon_{IB}{}^{CD}
E_C{}^{ \mu}  T^{B}{}_{ AD}
-\epsilon_{IB}{}^{ CD}
E_D{}^{ \mu} \omega^B{}_{ AC})+e^{-1}\partial_\nu\Big(eM^I\epsilon_{IA}{}^{CD}   E_C{}^{ \nu}E_D{}^{ \mu}\Big)\Big]\nonumber\\
&&-N^I(E_I{}^{ \mu} \omega^\rho{}_{ A \rho}	- \omega^\mu{}_{AI} -	T^\mu{}_{ AI}     	-	v_A	E_I{}^{ \mu})+e^{-1}\partial_\nu\Big(eN^I(E_A{}^{ \nu } E_I{}^{ \mu}  - E_A{}^{ \mu } E_I{}^{ \nu})\Big)\nonumber\\
&&-O^{IJK}H_{IJKA}{}^{\mu}+e^{-1}\partial_\nu\Big(eO^{IJK}L_{IJKA}{}^{\mu\nu}\Big)- \mathcal{L}_{\rm Tele}E_A{}^{\mu}+2E_A{}^{\nu}g^{\mu\alpha}\sum_{i=2}^{5}\mathcal{G}^{(i)}{}_{\alpha\nu}=2\kappa^2\Theta_A{}^{\mu}\,,
\end{eqnarray}
where the quantities $M^{I},\ N^I,\ O^{IJK}, \ H_{IJKA}{}^{\mu} $ and $L_{IJKA}{}^{\mu\nu}$ are given by \eqref{M}, \eqref{N}, \eqref{O}, \eqref{tensorial0} and \eqref{tensorial}, respectively. The terms $\mathcal{G}^{(i)}{}_{\alpha\nu}$  $\sum_{i=2}^{5}\mathcal{G}^{(i)}{}_{\mu\nu}$ were explicitly found in \cite{Capozziello:2018gms} (see Eqs.~(13a)-(13d) there). The complete derivation of the Teleparallel terms are written in the Appendix~\eqref{appendix1} for completeness. The quantity $S_A{}^{\lambda\mu}=\frac{1}{4}(T^{A}{}_{\lambda\mu}-T^{\lambda}{}_A{}^{\mu}-T^{\mu}{}_A{}^\lambda)+\frac{1}{2}(\delta^{\mu}{}_A T^\lambda-\delta^{\lambda}_A T^\mu)$ is the so-called superpotential.

Variations of the action with respect to the scalar field give us the modified Klein Gordon equation,
\begin{eqnarray}
\mathring{\nabla}^\mu\Big(J_{\mu\rm -Tele}+\sum_{i=2}^{5}J^{i}_\mu\Big)=P_{\phi-\rm Tele}+\sum_{i=2}^{5}P_{\phi}^i\,,
\end{eqnarray}
where $J_{\mu\rm -Tele}$ and $P_{\phi-\rm Tele}$ are defined as
\begin{eqnarray}
J_{\mu\rm -Tele}&=&-G_{\rm Tele,X}(\mathring{\nabla}_\mu\phi)+G_{\rm Tele,I_2} v_\mu+2G_{\rm Tele,J_1}a_\mu a^\nu\mathring{\nabla}_\nu \phi-G_{\rm Tele,J_3}v_\alpha t_{\mu}{}^{\nu\alpha}(\mathring{\nabla}_\nu\phi)\nonumber\\
&&-2G_{\rm Tele,J_5}t^{\beta\nu\alpha}t_{\beta\mu\alpha}(\mathring{\nabla}_\nu\phi)+2G_{\rm Tele,J_8}t^{\alpha\nu}{}_{\mu}t_{\alpha\nu}{}^{\beta}(\mathring{\nabla}_\beta\phi)-2G_{\rm Tele,J_6}t^{\nu\alpha\beta}t_{\mu}{}^{\sigma}{}_\nu(\mathring{\nabla}_\alpha\phi)(\mathring{\nabla}_\beta\phi)(\mathring{\nabla}_\sigma\phi)\,,\nonumber\\
&&-G_{\rm Tele,J_{10}} a^\nu (\mathring{\nabla}_\alpha \phi) (\epsilon^{\mu}{}_{\nu\rho\sigma}t^{\alpha\rho\sigma}+\epsilon^{\alpha}{}_{\nu\rho\sigma}t^{\mu\rho\sigma})\,,\label{Jtele2}\\
P_{\phi-\rm Tele}&=&G_{\rm Tele,\phi}\,.\label{PTele2}
\end{eqnarray}
For more details about the derivation of these equations, see the appendix~\S.\ref{sec:varphi}. Using $\mathring{R}=-T+B$, one finds that $P_{\phi}^i$ is given by~\cite{Capozziello:2018gms}
\begin{subequations}\label{Pphi}
	\begin{align}
	P_{\phi}^{2} &= G_{2,\phi}\,,\\
	P_{\phi}^{3} &= \mathring{\nabla}_{\mu}G_{3,\phi}\mathring{\nabla}^{\mu} \phi \,,\\
	P_{\phi}^{4} &= G_{4,\phi}(-T+B) + G_{4,\phi X} \left[ (\mathring{\square} \phi)^2 - (\mathring{\nabla}_{\mu}\mathring{\nabla}_{\nu}\phi)^2\right]\,,\\
	P_{\phi}^{5} &= -\mathring{\nabla}_{\mu}G_{5,\phi} \mathring{G}^{\mu\nu}\mathring{\nabla}_{\nu}\phi - \frac{1}{6}G_{5,\phi X}\left[(\square \phi)^3 - 3 \square \phi (\mathring{\nabla}_{\mu}\mathring{\nabla}_{\nu} \phi)^2 + 2 (\mathring{\nabla}_{\mu}\mathring{\nabla}_{\nu}\phi)^3 \right]\,,
	\end{align}
\end{subequations}
and  $J^{i}_\mu$ is defined as~\cite{Capozziello:2018gms}
\begin{subequations}\label{Jis}
	\begin{align}
	J_{\mu}^{2} &= -\mathcal{L}_{2,X}\mathring{\nabla}_{\mu}\phi \,,\\
	J_{\mu}^{3} &= -\mathcal{L}_{3,X}\mathring{\nabla}_{\mu}\phi + G_{3,X} \mathring{\nabla}_{\mu}X + 2 G_{3,\phi} \mathring{\nabla}_{\mu} \phi \,,\\
	J_{\mu}^{4} &= - \mathcal{L}_{4,X}\mathring{\nabla}_{\mu} \phi +2 G_{4,X}\mathring{R}_{\mu\nu}\mathring{\nabla}^{\nu} \phi  - 2 G_{4,XX}\left(\mathring{\square} \phi \mathring{\nabla}_{\mu}X - \mathring{\nabla}^{\nu} X \mathring{\nabla}_{\mu}\mathring{\nabla}_{\nu} \phi \right) \nonumber \\
	&-2 G_{4,\phi X} (\mathring{\square} \phi \mathring{\nabla}_{\mu}\phi + \mathring{\nabla}_{\mu}X) \,,\\
	J_{\mu}^{5} &= -\mathcal{L}_{5,X}\mathring{\nabla}_{\mu}\phi - 2 G_{5,\phi}\mathring{G}_{\mu\nu} \mathring{\nabla}^{\nu} \phi \nonumber \\
	&-G_{5,X}\left[ \mathring{G}_{\mu\nu}\mathring{\nabla}^{\nu} X + \mathring{R}_{\mu\nu}\square \phi \mathring{\nabla}^{\nu} \phi - \mathring{R}_{\nu \lambda} \mathring{\nabla}^{\nu} \phi \mathring{\nabla}^{\lambda}\mathring{\nabla}_{\mu} \phi - \mathring{R}_{\alpha \mu \beta \nu}\mathring{\nabla}^{\nu}\phi \mathring{\nabla}^{\alpha} \mathring{\nabla}^{\beta}\phi\right]  \nonumber \\
	&+G_{5,XX} \Big\{ \frac{1}{2}\mathring{\nabla}_{\mu}X \left[(\mathring{\square} \phi)^2 - (\mathring{\nabla}_{\alpha}\mathring{\nabla}_{\beta}\phi)^2 \right]- \mathring{\nabla}_{\nu}X\left(\mathring{\square} \phi \mathring{\nabla}_{\mu}\mathring{\nabla}^{\nu}\phi - \mathring{\nabla}_{\alpha}\mathring{\nabla}_{\mu}\phi\mathring{\nabla}^{\alpha}\mathring{\nabla}^{\nu}\phi\right)\Big\}  \nonumber \\
	&+G_{5,\phi X} \Big\{ \frac{1}{2}\mathring{\nabla}_{\mu}\phi \left[(\mathring{\square} \phi)^2 - (\mathring{\nabla}_{\alpha}\mathring{\nabla}_{\beta}\phi)^2 \right] + \mathring{\square} \phi \mathring{\nabla}_{\mu}X -\mathring{\nabla}^{\nu}X \mathring{\nabla}_{\nu}\mathring{\nabla}_{\mu}\phi  \Big\}\,.
	\end{align}
\end{subequations}
Note that no terms from the matter Lagrangian appear, since we are working in the Jordan conformal frame, in which there is no direct coupling term between the scalar and matter fields. To fully express all terms in the above equations in terms only depending on Teleparallel quantities, one can use the following identities
\begin{eqnarray}
\mathring{R}^{\lambda}\,_{\mu\sigma\nu} &=& \mathring{\nabla}_{\nu}K_{\sigma}{}^{\lambda}{}_{\mu} -
\mathring{\nabla}_{\sigma}K_{\nu}{}^{\lambda}{}_{\mu} +
K_{\sigma}{}^{\rho}{}_{\mu}K_{\nu}{}^{\lambda}{}_{\rho} -
K_{\sigma}{}^{\lambda}{}_{\rho}K_{\nu}{}^{\rho}{}_{\mu}\,,\\
\mathring{R}_{\mu\nu} &=& \mathring{\nabla}_{\nu}K_{\lambda}{}^{\lambda}{}_{\mu} -
\mathring{\nabla}_{\lambda}K_{\nu}{}^{\lambda}{}_{\mu} +
K_{\lambda}{}^{\rho}{}_{\mu}K_{\nu}{}^{\lambda}{}_{\rho} -
K_{\lambda}{}^{\lambda}{}_{\rho}K_{\nu}{}^{\rho}{}_{\mu}\,,\\
\mathring{G}_{\mu\nu}&=&e^{-1}e^{A}{}_{\mu}g_{\nu\rho}\partial_\sigma(e S_A{}^{\rho\sigma})-S_{B}{}^{\sigma}{}_{\nu}T^{B}{}_{\sigma\mu}+\frac{1}{4}T g_{\mu\nu}-e^{A}{}_\mu \omega ^{B}{}_{A\sigma}S_{B\nu}{}^{\sigma}\,.
\end{eqnarray}
Here, the term $K_{\mu}{}^{\lambda}{}_{\nu} =(1/2) (T^{\lambda}{}_{\mu\nu} - T_{\nu\mu}{}^{\lambda} +T_{\mu}{}^{\lambda}{}_{\nu})$ is the contortion tensor.

\section{Post-Newtonian limit in BDLS Theory}
\label{sec:PPN}
We now come to the post-Newtonian limit of the class of theories displayed in the previous section. In order to simplify the calculation and the result we display later, we first introduce a new parametrization of the action in section~\ref{ssec:parchange}. We then briefly review the post-Newtonian expansion of the field variables and post-Newtonian energy-momentum in section~\ref{ssec:ppnexp}. To apply this formalism to the theories at hand, we perform a Taylor expansion in section~\ref{ssec:taylor}. It turns out that we must impose that certain Taylor coefficients vanish; we list these restrictions in section~\ref{ssec:restrict}. In section~\ref{ssec:ppnsol}, we finally come to solve the field equations. The resulting PPN parameters are presented in section~\ref{ssec:ppnpar}.

\subsection{Change of parametrization}\label{ssec:parchange}
For the calculation of the post-Newtonian limit it is helpful to rewrite the teleparallel Lagrangian \(G_{\text{Tele}}\) in the equivalent form
\begin{equation}
F(\mathcal{T}_1, \mathcal{T}_2, \mathcal{T}_3, X, Y, \phi, \mathbb{J})\,,
\end{equation}
where we used the shorthand notation \(\mathbb{J} = (J_1, J_3, J_5, J_6, J_8, J_{10})\), and where we have introduced the new terms
\begin{equation}
\mathcal{T}_1 = T^{\mu\nu\rho}T_{\mu\nu\rho}\,, \quad
\mathcal{T}_2 = T^{\mu\nu\rho}T_{\rho\nu\mu}\,, \quad
\mathcal{T}_3 = T^\mu{}_{\mu\rho}T_\nu{}^{\nu\rho}\,, \quad
Y = g^{\mu\nu}T^{\rho}{}_{\rho\mu}\phi_{,\nu}\,.
\end{equation}
To see that this is simply an equivalent rewriting, note that these terms are related to the previously defined torsion terms by
\begin{equation}
T_{\text{ax}} = \frac{1}{18}(\mathcal{T}_1 - 2\mathcal{T}_2)\,, \quad
T_{\text{ten}} = \frac{1}{2}(\mathcal{T}_1 + \mathcal{T}_2 - \mathcal{T}_3)\,, \quad
T_{\text{vec}} = \mathcal{T}_3\,, \quad
T = \frac{1}{4}\mathcal{T}_1 + \frac{1}{2}\mathcal{T}_2 - \mathcal{T}_3\,, \quad
I_2 = Y\,,
\end{equation}
so that we can express \(G_{\text{Tele}}\) as
\begin{equation}
G_{\text{Tele}}(\phi, X, T, T_{\text{ax}}, T_{\text{vec}}, I_2, \mathbb{J}) = G_{\text{Tele}}\left(\phi, X, \frac{1}{4}\mathcal{T}_1 + \frac{1}{2}\mathcal{T}_2 - \mathcal{T}_3, \frac{1}{18}(\mathcal{T}_1 - 2\mathcal{T}_2), \mathcal{T}_3, Y, \mathbb{J}\right)\,.
\end{equation}
Conversely, we can express \(F\) in terms of the original variables as
\begin{equation}
F(\mathcal{T}_1, \mathcal{T}_2, \mathcal{T}_3, X, Y, \phi, \mathbb{J}) = F\left(2(T + T_{\text{vec}}) + 9T_{\text{ax}}, T + T_{\text{vec}} - \frac{9}{2}T_{\text{ax}}, T_{\text{vec}}, X, I_2, \phi, \mathbb{J}\right)\,.
\end{equation}
There are two reasons for this change of variables and rewriting of the Lagrangian. First, it turns out that the field equations and derivation of the post-Newtonian limit become simpler in the newly introduced variables. Second, these variables allow for a more direct comparison with previous results on the post-Newtonian limit of other theories, which now become obvious special cases of the class of theories we consider here~\cite{Hohmann:2015kra,Ualikhanova:2019ygl,Emtsova:2019qsl,Flathmann:2019khc}.

\subsection{Post-Newtonian expansion}\label{ssec:ppnexp}
We now come to a post-Newtonian approximation of the field equations of the BDLS theory detailed in the previous sections. Hereby we follow the parametrized post-Newtonian approach for teleparallel gravity theories developed in~\cite{Ualikhanova:2019ygl} and its adaptation for scalar-torsion theories used in~\cite{Emtsova:2019qsl,Flathmann:2019khc}. The starting point of this formalism is the assumption that the energy-momentum tensor of the source matter is given by that of a perfect fluid, and thus takes the form
\begin{equation}\label{eq:tmunu}
\Theta^{\mu\nu} = (\rho + \rho\Pi + p)u^{\mu}u^{\nu} + pg^{\mu\nu}
\end{equation}
with \(u^{\mu}u^{\nu}g_{\mu\nu} = -1\). Further, one assumes that the velocity \(v^i = u^i/u^0\) of the source matter is small, \(|\vec{v}| \ll c \equiv 1\), compared to the speed of light, in a particular frame of reference. Based on this assumption, one promotes the velocity to a perturbation parameter, and assigns velocity orders \(\mathcal{O}(n) \propto |\vec{v}|^n\) to all quantities. For the matter variables constituting the energy-momentum tensor~\eqref{eq:tmunu}, which are the rest mass density \(\rho\), specific internal energy \(\Pi\) and pressure \(p\), one assigns velocity orders \(\mathcal{O}(2)\) to \(\rho\) and \(\Pi\) and \(\mathcal{O}(4)\) to \(p\), taking into account their orders of magnitude in the Solar System; see~\cite{Will:1993ns} for their definition and more thorough explanation of their properties. One further assumes that the gravitational field is quasi-static, which means that it changes only following the motion of the source matter. Hence, time derivatives are weighted with an additional velocity order \(\partial_0 \sim \mathcal{O}(1)\). Finally, one assumes that the background of the perturbative expansion is given by a diagonal tetrad and a constant value of the scalar field,
\begin{equation}\label{eq:background}
\order{e}{0}^A{}_{\mu} = \Delta^A{}_\mu = \mathrm{diag}(1,1,1,1)\,, \quad
\order{\phi}{0} = \Phi\,.
\end{equation}
One then performs a perturbative expansion of the tetrad and the scalar field of the form
\begin{equation}
e^A{}_{\mu} = \sum_k\order{e}{k}^A{}_{\mu}\,, \quad
\phi = \sum_k\order{\phi}{k}\,.
\end{equation}
Note that we do not introduce an expansion of the spin connection here, since we assume the Weitzenböck gauge \(\omega^A{}_{B\mu} \equiv 0\) at all perturbation orders. Further, it is useful to lower the Lorentz index of the tetrad perturbations using the Minkowski metric and to convert it to a spacetime index using the background tetrad, hence defining
\begin{equation}\label{eq:tetlowdef}
\order{e}{k}_{\mu\nu} = \Delta^A{}_{\mu}\eta_{AB}\order{e}{k}^B{}_{\nu}\,.
\end{equation}
Finally, we perform a $3+1$ split of the tetrad components into temporal and spatial parts. Following the procedure detailed in~\cite{Ualikhanova:2019ygl,Emtsova:2019qsl,Flathmann:2019khc}, we find that the only relevant and non-vanishing components of the field perturbations are given by
\begin{equation}\label{eq:ppnfields}
\order{e}{2}_{00}\,, \quad
\order{e}{2}_{ij}\,, \quad
\order{e}{3}_{0i}\,, \quad
\order{e}{3}_{i0}\,, \quad
\order{e}{4}_{00}\,, \quad
\order{\phi}{2}\,, \quad
\order{\phi}{4}\,.
\end{equation}
These are the components we will solve for in order to determine the post-Newtonian limit and hence the PPN parameters.

\subsection{Taylor expansion of parameter functions}\label{ssec:taylor}
In order to perform the perturbative expansion of the field equations around the background (vacuum) solution~\eqref{eq:background}, we must perform a perturbative expansion of the parameter functions which appear in the action around the same background. For the parameter functions \(G_2, G_3, G_4, G_5, G_{\text{Tele}}, F\) this means that we must perform a Taylor expansion around \(\phi = \Phi\), and all other arguments to these functions vanish. For the Horndeski part, this expansion takes the form
\begin{equation}
\begin{split}
G_i(\phi, X) &= G_i(\Phi, 0) + G_{i,\phi}(\Phi, 0)\psi + G_{i,X}(\Phi, 0)X + \frac{1}{2}G_{i,\phi\phi}(\Phi, 0)\psi^2 + G_{i,\phi X}(\Phi, 0)\psi X + \frac{1}{2}G_{i,XX}(\Phi, 0)X^2 + \ldots\\
&= \GG_i + \GG_{i,\phi}\psi + \GG_{i,X}X + \frac{1}{2}\GG_{i,\phi\phi}\psi^2 + \GG_{i,\phi X}\psi X + \frac{1}{2}\GG_{i,XX}X^2 + \ldots
\end{split}
\end{equation}
for \(i = 2,3,4,5\), where we introduced boldface letters to denote the constant Taylor coefficients at the background level. We have decomposed the scalar field as $\phi=\Phi+\psi$, where $\Phi=\order{\phi}{0}$ is the scalar field evaluated at the background and $\psi=\sum_{i=1}^{4}\order{\phi}{i}$ is the sum of all the order perturbations. Similarly, we introduce the notation
\begin{equation}
\GG_{\text{Tele}}, \GG_{\text{Tele},\phi}, \GG_{\text{Tele},X}, \GG_{\text{Tele},T}, \GG_{\text{Tele},T_{\text{ax}}}, \GG_{\text{Tele},T_{\text{vec}}}, \GG_{\text{Tele},I_2}
\end{equation}
for the background value of \(G_{\text{Tele}}\) and its derivatives, as well as
\begin{equation}
\FF, \FF_{,1}, \FF_{,2}, \FF_{,3}, \FF_{,X}, \FF_{,Y}, \FF_{\phi}
\end{equation}
for the background value of \(F\) and its derivatives. Note that we do not introduce any notation for the derivatives with respect to \(\mathbb{J}\), as it will turn out that these do not enter the field equations at the post-Newtonian order we consider. For the relevant Taylor coefficients, we find the relations
\begin{align}
\FF_{,1} &= \frac{9\GG_{\text{Tele},T} + 2\GG_{\text{Tele},T_{\text{ax}}}}{36}\,, &
\FF_{,2} &= \frac{9\GG_{\text{Tele},T} - 2\GG_{\text{Tele},T_{\text{ax}}}}{18}\,, &
\FF_{,3} &= \GG_{\text{Tele},T_{\text{vec}}} - \GG_{\text{Tele},T}\,,\\
\FF_{,\phi 1} &= \frac{9\GG_{\text{Tele},\phi T} + 2\GG_{\text{Tele},\phi T_{\text{ax}}}}{36}\,, &
\FF_{,\phi 2} &= \frac{9\GG_{\text{Tele},\phi T} - 2\GG_{\text{Tele},\phi T_{\text{ax}}}}{18}\,, &
\FF_{,\phi 3} &= \GG_{\text{Tele},\phi T_{\text{vec}}} - \GG_{\text{Tele},\phi T}\,,
\end{align}
as well as their inverses
\begin{align}
\GG_{\text{Tele},T} &= 2\FF_{,1} + \FF_{,2}\,, &
\GG_{\text{Tele},T_{\text{vec}}} &= 2\FF_{,1} + \FF_{,2} + \FF_{,3}\,, &
\GG_{\text{Tele},T_{\text{ax}}} &= \frac{18\FF_{,1} - 9\FF_{,2}}{2}\,,\\
\GG_{\text{Tele},\phi T} &= 2\FF_{,\phi 1} + \FF_{,\phi 2}\,, &
\GG_{\text{Tele},\phi T_{\text{vec}}} &= 2\FF_{,\phi 1} + \FF_{,\phi 2} + \FF_{,\phi 3}\,, &
\GG_{\text{Tele},\phi T_{\text{ax}}} &= \frac{18\FF_{,\phi 1} - 9\FF_{,\phi 2}}{2}
\end{align}
for the derivatives with respect to the various torsion scalars, while derivatives with respect to the scalar field terms \(\phi, X, Y = I_2\) take the same form in both parametrizations.

\subsection{Restrictions on the considered theories}\label{ssec:restrict}
In order to be able to solve the post-Newtonian field equations and express the solution in terms of the PPN potentials, we must impose several restrictions on the parameter functions. The first restriction is imposed by the background field equations. We find that our assumed background~\eqref{eq:background} is a solution to the field equations only if we assume
\begin{equation}\label{eq:vacuum}
\FF + \GG_2 = 0\,, \quad
\FF_{,\phi} + \GG_{2,\phi} = 0\,,
\end{equation}
or analogously in terms of \(G_{\text{Tele}}\),
\begin{equation}
\GG_{\text{Tele}} + \GG_2 = 0\,, \quad
\GG_{\text{Tele},\phi} + \GG_{2,\phi} = 0\,,
\end{equation}
Further, we restrict ourselves to theories in which the scalar field is massless. Without this restriction, Yukawa-like terms would appear, which would require an extension of the standard PPN formalism~\cite{Zaglauer:1990yh,Helbig:1991pk}. Hence, we set
\begin{equation}\label{eq:massless}
\FF_{,\phi\phi} + \GG_{2,\phi\phi} = 0\,, \quad
\FF_{,\phi\phi\phi} + \GG_{2,\phi\phi\phi} = 0\,,
\end{equation}
which we can also write as
\begin{equation}
\GG_{\text{Tele},\phi\phi} + \GG_{2,\phi\phi} = 0\,, \quad
\GG_{\text{Tele},\phi\phi\phi} + \GG_{2,\phi\phi\phi} = 0\,.
\end{equation}
Finally, we remove terms which are of higher than second derivative order, and would thus lead to terms in the solution involving higher derivatives of the source terms~\cite{Gladchenko:1990nw}. These terms are eliminated by~\cite{Hohmann:2015kra}
\begin{equation}\label{eq:higherorder}
\GG_{3,X} - 3\GG_{4,\phi X} = 0\,, \quad
\GG_{4,X} - \GG_{5,\phi} = 0\,.
\end{equation}
In the following, we will consider only theories which satisfy these restrictions.

\subsection{Post-Newtonian solution}\label{ssec:ppnsol}
We are now in the position to derive a perturbative solution to the field equations. We proceed in three steps with increasing velocity (perturbation) order. We solve the field equations at \(\mathcal{O}(2)\) in section~\ref{sssec:ppnsol2}, at \(\mathcal{O}(3)\) in section~\ref{sssec:ppnsol3} and at \(\mathcal{O}(4)\) in section~\ref{sssec:ppnsol4}. See, e.g., \cite{Will:1993ns} for the definition of the post-Newtonian potentials \(U, V_i, W_i, \Phi_1, \Phi_2, \Phi_3, \Phi_4, \Phi_W, \mathfrak{A}, \mathfrak{B}\) we use in this section.

\subsubsection{Second velocity order}\label{sssec:ppnsol2}
Since we have already solved the zeroth-order (vacuum) field equations by restricting the Taylor coefficients to the values~\eqref{eq:vacuum}, we continue by solving the field equations at the second velocity order. These are given by the temporal tetrad equation
\begin{equation}
(2\FF_{,1} + \FF_{,2} + \FF_{,3})\bar{\triangle}\order{e}{2}_{00} + (\FF_{,3} + \GG_4)(\order{e}{2}_{ij,ij} - \bar{\triangle}\order{e}{2}_{ii}) + \frac{1}{2}(\FF_{,Y} - 2\GG_{4,\phi})\bar{\triangle}\order{\phi}{2} = \kappa^2\rho\,,
\end{equation}
where \(\bar{\triangle} = \delta^{ij}\partial_i\partial_j\) denotes the Laplacian of the spatial background metric, the spatial tetrad equation
\begin{multline}
\frac{1}{2}(2\FF_{,2} - \GG_4)\bar{\triangle}(\order{e}{2}_{ij} - \order{e}{2}_{kj,ik} - \order{e}{2}_{ik,jk}) + \frac{1}{2}(4\FF_{,2} - \GG_4)(\bar{\triangle}\order{e}{2}_{ji} - \order{e}{2}_{jk,ik}) + (\FF_{,3} + \GG_4)(\order{e}{2}_{00,ij} - \order{e}{2}_{kk,ij}) + \frac{1}{2}(2\FF_{,2} + 2\FF_{,3} + \GG_4)\order{e}{2}_{ki,jk}\\
+ \frac{1}{2}(\FF_{,Y} - 2\GG_{4,\phi})\order{\phi}{2}_{,ij} + \left[(\FF_{,3} + \GG_4)(\bar{\triangle}\order{e}{2}_{ii} - \order{e}{2}_{ij,ij} - \bar{\triangle}\order{e}{2}_{00}) - \frac{1}{2}(\FF_{,Y} - 2\GG_{4,\phi})\bar{\triangle}\order{\phi}{2}\right]\delta_{ij} = 0
\end{multline}
and the scalar field equation
\begin{equation}
(\FF_{,Y} - 2\GG_{4,\phi})(\bar{\triangle}\order{e}{2}_{00} - \bar{\triangle}\order{e}{2}_{ii} + \order{e}{2}_{ij,ij}) - (\FF_{,X} + \GG_{2,X} - 2\GG_{3,\phi})\bar{\triangle}\order{\phi}{2} = 0\,.
\end{equation}
In order to solve these equations, we make an ansatz of the form
\begin{equation}
\order{e}{2}_{00} = a_1U\,, \quad
\order{e}{2}_{ij} = a_2U\delta_{ij}\,, \quad
\order{\phi}{2} = a_3U\,,
\end{equation}
where \(a_{1,2,3}\) are unknown constants which are to be determined from the field equations. This ansatz is motivated by the structure of the field equations, which contain second-order spatial derivatives of the field variables, which must be proportional to the matter density, and can be obtained, e.g., from a gauge-invariant formalism as shown in~\cite{Hohmann:2019qgo}. Inserting this ansatz into the second order field equations yields the solution
\begin{subequations}
\begin{align}
a_1 &= -\kappa^2\frac{(2\FF_{,1} + \FF_{,2} + 2\FF_{,3} + \GG_4)(\FF_{,X} + \GG_{2,X} - 2\GG_{3,\phi}) + (\FF_{,Y} - 2\GG_{4,\phi})^2}{2\pi(2\FF_{,1} + \FF_{,2} - \GG_4)[2(2\FF_{,1} + \FF_{,2} + 3\FF_{,3} + 2\GG_4)(\FF_{,X} + \GG_{2,X} - 2\GG_{3,\phi}) + 3(\FF_{,Y} - 2\GG_{4,\phi})^2]}\,,\\
a_2 &= -\kappa^2\frac{2(\FF_{,3} + \GG_4)(\FF_{,X} + \GG_{2,X} - 2\GG_{3,\phi}) + (\FF_{,Y} - 2\GG_{4,\phi})^2}{4\pi(2\FF_{,1} + \FF_{,2} - \GG_4)[2(2\FF_{,1} + \FF_{,2} + 3\FF_{,3} + 2\GG_4)(\FF_{,X} + \GG_{2,X} - 2\GG_{3,\phi}) + 3(\FF_{,Y} - 2\GG_{4,\phi})^2]}\,,\\
a_3 &= -\kappa^2\frac{(\FF_{,Y} - 2\GG_{4,\phi})}{2\pi[2(2\FF_{,1} + \FF_{,2} + 3\FF_{,3} + 2\GG_4)(\FF_{,X} + \GG_{2,X} - 2\GG_{3,\phi}) + 3(\FF_{,Y} - 2\GG_{4,\phi})^2]}\,.
\end{align}
\end{subequations}
We will make use of this solution in the remainder of the calculation.

\subsubsection{Third velocity order}\label{sssec:ppnsol3}
We then proceed with the third velocity order. The corresponding field equations read
\begin{multline}
\frac{1}{2}(4\FF_{,1} - \GG_4)(\bar{\triangle}\order{e}{3}_{i0} - \order{e}{2}_{ij,0j}) + \frac{1}{2}(2\FF_{,2} - \GG_4)(\bar{\triangle}\order{e}{3}_{0i} - \order{e}{3}_{0j,ij} - \order{e}{2}_{ji,0j})\\
+ \frac{1}{2}(2\FF_{,2} + 2\FF_{,3} + \GG_4)\order{e}{3}_{j0,ij} - (\FF_{,3} + \GG_4)\order{e}{2}_{jj,0i} + \frac{1}{2}(\FF_{,Y} - 2\GG_{4,\phi}) = -\kappa^2\rho v_i
\end{multline}
and
\begin{multline}
\frac{1}{2}(2\FF_{,2} - \GG_4)(\bar{\triangle}\order{e}{3}_{i0} - \order{e}{3}_{j0,ij} - \order{e}{2}_{ij,0j}) + \frac{1}{2}(4\FF_{,1} - \GG_4)(\bar{\triangle}\order{e}{3}_{0i} - \order{e}{3}_{0j,ij})\\
+ \frac{1}{2}(2\FF_{,2} + 2\FF_{,3} + \GG_4)\order{e}{2}_{ji,0j} - (\FF_{,3} + \GG_4)\order{e}{2}_{jj,0i} + (2\FF_{,1} + \FF_{,2} + \FF_{,3})\order{e}{2}_{00,i0} + \frac{1}{2}(\FF_{,Y} - 2\GG_{4,\phi}) = -\kappa^2\rho v_i\,.
\end{multline}
Solving these equations using an ansatz of the form
\begin{equation}
\order{e}{3}_{i0} = b_1V_i + b_2W_i\,, \quad
\order{e}{3}_{0i} = b_3V_i + b_4W_i\,,
\end{equation}
which follows from a similar motivation as for the second order field equations, we find that the solution is not unique. This is a consequence of the diffeomorphism invariance of the theory and the resulting gauge invariance of the post-Newtonian approximation. In order to solve the equations, one therefore needs to fix a gauge. Here we choose to work in the PPN gauge, which is determined only after the fourth velocity order is solved~\cite{Ualikhanova:2019ygl,Emtsova:2019qsl,Flathmann:2019khc}. At the third velocity order we thus obtain the partial solution
\begin{equation}
b_1 = b_3 + b_4 = \frac{\kappa^2}{4\pi(2\FF_{,1} + \FF_{,2} - \GG_4)}\,, \quad
b_2 = 0\,,
\end{equation}
while the difference \(b_3 - b_4\) is left to be determined in the following step.

\subsubsection{Fourth velocity order}\label{sssec:ppnsol4}
In the last step, we come to solve the fourth velocity order of the perturbative expansion of the field equations. At this perturbation order the equations become very lengthy, and so we will not display them here, but only sketch the procedure. The equations we must solve at the fourth velocity order are given by the temporal and spatial tetrad field equations, as well as the scalar field equations. It turns out that by taking a suitable linear combination of these equations, the unknown fourth-order terms
\begin{equation}
\bar{\triangle}\order{e}{4}_{ii}\,, \quad
\order{e}{4}_{ij,ij}\,, \quad
\bar{\triangle}\order{\phi}{4}\,,
\end{equation}
which are not relevant for determining the post-Newtonian parameters, drop out of the equations, and the only remaining fourth-order term we need to solve for is \(\bar{\triangle}\order{e}{4}_{00}\). This is achieved by making an ansatz of the form
\begin{equation}
\order{e}{4}_{00} = c_1\Phi_1 + c_2\Phi_2 + c_3\Phi_3 + c_4\Phi_4 + c_5U^2\,,
\end{equation}
which follows from the fact that its second-order derivatives reproduce the matter source terms in the corresponding field equations at the fourth velocity order. We can then solve the field equations for the constants \(c_1, \ldots, c_5\), as well as the so far undetermined difference \(b_3 - b_4\) left from the third velocity order. This solution then determines all tetrad components which we need in order to obtain the full set of PPN parameters.

\subsection{PPN parameters}\label{ssec:ppnpar}
Finally, we can compare our result for the post-Newtonian tetrad solution to the standard PPN form, according to which its symmetric components are given by~\cite{Hohmann:2019qgo}
\begin{subequations}\label{eq:standardppnt}
\begin{align}
\order{e}{2}_{00} &= U\,,\\
\order{e}{2}_{(ij)} &= \gamma U\delta_{ij}\,,\\
\order{e}{3}_{(0i)} &= -\frac{1}{4}(3 + 4\gamma + \alpha_1 - \alpha_2 + \zeta_1 - 2\xi)V_i - \frac{1}{4}(1 + \alpha_2 - \zeta_1 + 2\xi)W_i\,,\\
\order{e}{4}_{00} &= \frac{1}{2}(1 - 2\beta)U^2 + \frac{1}{2}(2 + 2\gamma + \alpha_3 + \zeta_1 - 2\xi)\Phi_1 + (1 + 3\gamma - 2\beta + \zeta_2 + \xi)\Phi_2\nonumber\\
&\phantom{=}+ (1 + \zeta_3)\Phi_3 + (3\gamma + 3\zeta_4 - 2\xi)\Phi_4 - \xi\Phi_W - \frac{1}{2}(\zeta_1 - 2\xi)\mathfrak{A}\,.
\end{align}
\end{subequations}
By comparison with our result we find the PPN parameters
\begin{equation}
\xi = \alpha_1 = \alpha_2 = \alpha_3 = \zeta_1 = \zeta_2 = \zeta_3 =\zeta_4 = 0\,,
\end{equation}
which indicates that the class of theories we study is fully conservative, i.e., it does not exhibit any preferred-frame or preferred-location effects, or a violation of the conservation of total energy-momentum. The only PPN parameters, for which we obtain a deviation from their general relativity values, are \(\gamma\) and \(\beta\). For the former we find
\begin{equation}\label{eq:gammaefhu}
\gamma = 1 - \frac{(\FF_{,Y} - 2\GG_{4,\phi})^2 + 2(2\FF_{,1} + \FF_{,2} + \FF_{,3})(\FF_{,X} + \GG_{2,X} - 2\GG_{3,\phi})}{2(\FF_{,Y} - 2\GG_{4,\phi})^2 + 2(2\FF_{,1} + \FF_{,2} + 2\FF_{,3} + \GG_4)(\FF_{,X} + \GG_{2,X} - 2\GG_{3,\phi})}\,,
\end{equation}
which we can express in the alternative parametrization as
\begin{equation}\label{eq:gammabdls}
\gamma = 1 - \frac{(\GG_{\text{Tele},I_2} - 2\GG_{4,\phi})^2 + 2\GG_{\text{Tele},T_{\text{vec}}}(\GG_{\text{Tele},X} + \GG_{2,X} - 2\GG_{3,\phi})}{2(\GG_{\text{Tele},I_2} - 2\GG_{4,\phi})^2 + 2(2\GG_{\text{Tele},T_{\text{vec}}} - \GG_{\text{Tele},T} + \GG_4)(\GG_{\text{Tele},X} + \GG_{2,X} - 2\GG_{3,\phi})}\,.
\end{equation}
The result for \(\beta\), however, is very lengthy. It is possible to express the PPN parameters in a neat form by introducing the following combinations,
\begin{eqnarray}
\HH_{,1}&:=&\GG_{2,X}-2 \GG_{3,\phi}+\FF_{,X}=\GG_{2,X}-2 \GG_{3,\phi}+\GG_{\text{Tele},X}\,,\label{H1}\\
\HH_{,2}&:=&2 \FF_{\phi 1}+\FF_{\phi 2}=\GG_{\text{Tele},\phi T}\,,\\
\HH_{,3}&:=&\FF_{,Y}-2 \GG_{4,\phi}=\GG_{\text{Tele},I_2}-2 \GG_{4,\phi}\,,\\
\HH_{,4}&:=&2 \FF_{,1}+\FF_{,2}+\FF_{,3}=\GG_{\text{Tele},T_{\text{vec}}}\,,\\
\HH_{,5}&:=&2 \FF_{,1}+\FF_{,2}+2 \FF_{,3}+\GG_4= 2 \GG_{\text{Tele},T_{\text{vec}}}-\GG_{\text{Tele},T}+\GG_4\,,\label{H5}\\
\HH_{,6}&:=&\HH_{,3} \left(4 \HH_{,1} \GG_{4,\phi\phi}-2 \HH_{,1} \FF_{,\phi Y}+\HH_{,3} \left(\GG_{2,\phi X}-2 \GG_{3,\phi\phi}+\FF_{,\phi X}\right)\right)-2 \HH_{,1}^2 \FF_{\phi 3}\label{H6B}\\
&=&2 \Big(\GG_{2,X}-2 \GG_{3,\phi}+\GG_{\text{Tele},X}\Big) \Big(\left(\GG_{\text{Tele},\phi T}-\GG_{\text{Tele},\phi T_{\text{vec}}}\right) \left(\GG_{2,X}-2 \GG_{3,\phi}+\GG_{\text{Tele},X}\right)\nonumber\\
&&+\left(\GG_{\text{Tele},I_2}-2 \GG_{4,\phi}\right) \left(2 \GG_{4,\phi\phi}-\GG_{\text{Tele},\phi I_2}\right)\Big)+\left(\GG_{\text{Tele},I_2}-2 \GG_{4,\phi}\right){}^2 \left(\GG_{2,\phi X}-2 \GG_{3,\phi\phi}+\GG_{\text{Tele},\phi X}\right)\,,\label{H6}
\end{eqnarray}
giving us the following form of $\gamma$
\begin{eqnarray}\label{eq:gammapar}
\gamma&=&1-\frac{2 \HH_{,1} \HH_{,4}+\HH_{,3}^2}{2 \left(\HH_{,1} \HH_{,5}+\HH_{,3}^2\right)}\,,
\end{eqnarray}
and $\beta$ becomes
\begin{eqnarray}\label{eq:betapar}
\beta&=&1-\frac{\tilde{\beta}}{8 \left(\HH_{,1} \HH_{,5}+\HH_{,3}^2\right){}^2 \left(3 \HH_{,3}^2-2 \HH_{,1} \left(\HH_{,4}-2 \HH_{,5}\right)\right)}\,,
\end{eqnarray}
where the function $\tilde{\beta}$ is
\begin{eqnarray}
\tilde{\beta}&=&4 \HH_{,1}^2 \HH_{,3} \Big(\HH_{,2} \left(4 \HH_{,4} \HH_{,5}-2 \HH_{,4}^2-3 \HH_{,5}^2\right)+\HH_{,3} \left(3 \HH_{,4} \HH_{,5}-\HH_{,4}^2+\HH_{,5}^2\right)\Big)-4 \HH_{,1}^3 \HH_{,4} \HH_{,5} \left(\HH_{,4}-2 \HH_{,5}\right)\nonumber\\
&&+\HH_{,1} \HH_{,3}^3 \Big(8 \HH_{,2} \left(\HH_{,4}-2 \HH_{,5}\right)+\HH_{,3} \left(4 \HH_{,4}+7 \HH_{,5}\right)\Big)+\HH_{,3} \left(-6 \HH_{,2} \HH_{,3}^4+3 \HH_{,3}^5+2 \HH_{,6} \left(\HH_{,5}-2 \HH_{,4}\right){}^2\right)\nonumber\\
&&+\GG_{4,\phi} \Big(-8 \HH_{,1} \HH_{,3}^3 \left(\HH_{,4}-2 \HH_{,5}\right)+8 \HH_{,1}^2 \HH_{,3} \left(\HH_{,5}^2-\HH_{,4}^2\right)+6 \HH_{,3}^5\Big)\,.\label{betafinal}
\end{eqnarray}
Note the appearance of particular combinations of the Taylor coefficients in the denominators of the PPN parameters~\eqref{eq:gammapar} and~\eqref{eq:betapar}. If any of these denominators vanish, the PPN parameters diverge, and the theory is ill-defined. Indeed, one finds that these values correspond to cases in which either the scalar field or the metric becomes strongly coupled around the background vacuum solution we consider. For
\begin{equation}
\HH_{,1}\HH_{,5} + \HH_{,3}^2 \to 0
\end{equation}
the kinetic term of the metric in the Einstein frame vanishes, while for
\begin{equation}
3\HH_{,3}^2 - 2\HH_{,1}\left(\HH_{,4} - 2\HH_{,5}\right) \to 0
\end{equation}
the same happens for the kinetic term of the scalar field.

\section{Special cases}\label{sec:special}
In the previous section we have derived the PPN parameters of the most general class of BDLS teleparallel Horndeski theories. We now highlight a few special cases, which are of particular interest for various reasons. First, in section~\ref{ssec:puretele}, we consider the ``purely teleparallel'' class of theories, in which only the term \(G_{\text{Tele}}\) is present in the action. We then reproduce a number of previously obtained results for three well-known classes of theories: Horndeski gravity in section~\ref{ssec:horndeski}, scalar-torsion gravity in section~\ref{ssec:scaltors} and \(F(\mathcal{T}_1, \mathcal{T}_2, \mathcal{T}_3)\) gravity in section~\ref{ssec:fttt}. The general relativity limit of BDLS theory, which can be obtained in different ways, is discussed in section~\ref{ssec:grlimit}. Finally, we derive the most general class of theories with PPN parameters identical to those of general relativity in section~\ref{ssec:grmimic}; theories of this latter class are indistinguishable from general relativity at the level of their PPN parameters.

\subsection{Pure $G_{\text{Tele}}$ theories}\label{ssec:puretele}
A particular case is given if we consider theories in which only the teleparallel term \(G_{\text{Tele}}\) is present, while the usual Horndeski terms vanish, \(G_2 = G_3 = G_4 = G_5 = 0\). In this case we find the PPN parameters
\begin{equation}
\gamma = \frac{1}{2}\frac{\FF_{,Y}^2 + 2\FF_{,3}\FF_{,X}}{\FF_{,Y}^2 + \FF_{,X}(2\FF_{,1} + \FF_{,2} + 2\FF_{,3})}\,.
\end{equation}
These can alternatively be written as
\begin{equation}
\gamma = 1 - \frac{1}{2}\frac{\GG_{\text{Tele},I_2}^2 + 2\GG_{\text{Tele},T_{\text{vec}}}\GG_{\text{Tele},X}}{\GG_{\text{Tele},I_2}^2 + \GG_{\text{Tele},X}(2\GG_{\text{Tele},T_{\text{vec}}} - \GG_{\text{Tele},T})}\,,
\end{equation}
and in terms of $\HH$, this quantity reads as in~\eqref{eq:gammapar}.
For the $\beta$ parameter, the expressions look cumbersome in $\GG_{\text{Tele}}$ and $\FF$, but using the functions $\HH$ defined in~\eqref{H1}-\eqref{H5}, one finds the following neat expression,
\begin{eqnarray}
\beta&=&1-\frac{1}{8 \left(\HH_{,1} \HH_{,5}+\HH_{,3}^2\right){}^2 \left(3 \HH_{,3}^2-2 \HH_{,1} \left(\HH_{,4}-2 \HH_{,5}\right)\right)}\Big[-2 \HH_{,2} \Big(-4 \HH_{,1} \HH_{,3}^3 \left(\HH_{,4}-2 \HH_{,5}\right)\nonumber\\
&&+2 \HH_{,1}^2 \HH_{,3} \left(-4 \HH_{,4} \HH_{,5}+2 \HH_{,4}^2+3 \HH_{,5}^2\right)+3 \HH_{,3}^5\Big)+\HH_{,1} \HH_{,3}^4 \left(4 \HH_{,4}+7 \HH_{,5}\right)+4 \HH_{,1}^2 \HH_{,3}^2 \left(3 \HH_{,4} \HH_{,5}-\HH_{,4}^2+\HH_{,5}^2\right)\nonumber\\
&&-4 \HH_{,1}^3 \HH_{,4} \HH_{,5} \left(\HH_{,4}-2 \HH_{,5}\right)+2 \HH_{,3} \HH_{,6} \left(\HH_{,5}-2 \HH_{,4}\right){}^2+3 \HH_{,3}^6\Big]\,.
\end{eqnarray}
\subsection{Horndeski gravity}\label{ssec:horndeski}
The opposite case, compared to the previous one, is the well-known case of Horndeski gravity, which is obtained for \(G_{\text{Tele}} = 0\). In this case the PPN parameters reduce to
\begin{equation}
\gamma = 1 - \frac{4\GG_{4,\phi}^2}{\GG_4(\GG_{2,X} - 2\GG_{3,\phi}) + 4\GG_{4,\phi}^2}
\end{equation}
and
\begin{equation}
\beta = 1 + \frac{\GG_4\GG_{4,\phi}^2[\GG_4\GG_{4,\phi}(\GG_{2\phi X} - 2\GG_{3,\phi\phi}) + (\GG_{2,X} - 2\GG_{3,\phi})(\GG_{4,\phi}^2 - 2\GG_4\GG_{4,\phi\phi})]}{2[\GG_4(\GG_{2,X} - 2\GG_{3,\phi}) + 3\GG_{4,\phi}^2][\GG_4(\GG_{2,X} - 2\GG_{3,\phi}) + 4\GG_{4,\phi}^2]^2}\,.
\end{equation}
This result agrees with the PPN parameters found in~\cite{Hohmann:2015kra}.

\subsection{Scalar-torsion gravity}\label{ssec:scaltors}
A particular subclass of the pure teleparallel class discussed in section~\ref{ssec:puretele} is obtained if the torsion enters the action only through the linear combination
\begin{equation}
T = \frac{1}{4}\mathcal{T}_1 + \frac{1}{2}\mathcal{T}_2 - \mathcal{T}_3\,.
\end{equation}
In this case the parameter function reduces to
\begin{equation}
F(\mathcal{T}_1, \mathcal{T}_2, \mathcal{T}_3, X, Y, \phi, \mathbb{J}) = L\left(\frac{1}{4}\mathcal{T}_1 + \frac{1}{2}\mathcal{T}_2 - \mathcal{T}_3, X, Y, \phi\right)\,,
\end{equation}
where we also omitted the dependence on \(\mathbb{J}\), as it does not contribute to the post-Newtonian limit. Equivalently, we have
\begin{equation} \label{bb{L}}
G_{\text{Tele}}(\phi, X, T, T_{\text{ax}}, T_{\text{vec}}, I_2, \mathbb{J}) = L(T, X, I_2, \phi)\,.
\end{equation}
For this action, which was proposed in~\cite{Hohmann:2018dqh}, we find the PPN parameters
\begin{equation}
\gamma = 1 + \frac{\LL_{,Y}^2}{2\LL_{,T}\LL_{,X}-2\LL_{,Y}^2}
\end{equation}
and
\begin{multline}
\beta = 1 +\\
\frac{\LL_{,Y}\left[\LL_{,T}\LL_{,X}\LL_{,Y}^2\left(16\LL_{\phi T}-7\LL_{,Y}\right)+3\LL_{,Y}^4\left(\LL_{,Y}-2\LL_{\phi T}\right)-8\LL_{,T}^2\LL_{,X}^2\LL_{\phi T}+2\LL_{,T}^2\LL_{,Y}\left(2\LL_{,X}^2+\LL_{,Y}\LL_{\phi X}-2\LL_{,X}\LL_{\phi Y}\right)\right]}{8\left(4\LL_{,T}\LL_{,X}-3\LL_{,Y}^2\right)\left(\LL_{,Y}^2-\LL_{,T}\LL_{,X}\right)^2}\,,
\end{multline}
in agreement with an earlier result~\cite{Flathmann:2019khc}; where $\mathbf{L}$ is the Taylor series coefficients of the Lagrangian \eqref{bb{L}} evaluated at the background level.

\subsection{$F(\mathcal{T}_1, \mathcal{T}_2, \mathcal{T}_3)$ theories}\label{ssec:fttt}
Another special subclass of the pure teleparallel case is obtained if the function \(F\) does not depend on the scalar field, and hence takes the form
\begin{equation}
F(\mathcal{T}_1, \mathcal{T}_2, \mathcal{T}_3, X, Y, \phi, \mathbb{J}) = F(\mathcal{T}_1, \mathcal{T}_2, \mathcal{T}_3)\,.
\end{equation}
This action was studied in~\cite{Bahamonde:2017wwk,Hohmann:2018xnb}. In this case we find the PPN parameters
\begin{equation}
\gamma = 1 - \frac{2\FF_{,1} + \FF_{,2} + \FF_{,3}}{2\FF_{,1} + \FF_{,2} + 2\FF_{,3}}\,, \quad
\beta = 1 - \frac{1}{4}\frac{2\FF_{,1} + \FF_{,2} + \FF_{,3}}{2\FF_{,1} + \FF_{,2} + 2\FF_{,3}}\,.
\end{equation}
This is the result obtained in~\cite{Ualikhanova:2019ygl}. It may equivalently be recast in the form
\begin{equation}
\gamma = 1 - \frac{\GG_{\text{Tele},T_{\text{vec}}}}{2\GG_{\text{Tele},T_{\text{vec}}} - \GG_{\text{Tele},T}}\,, \quad
\beta = 1 - \frac{1}{4}\frac{\GG_{\text{Tele},T_{\text{vec}}}}{2\GG_{\text{Tele},T_{\text{vec}}} - \GG_{\text{Tele},T}}\,,
\end{equation}
using the parametrization through \(G_{\text{Tele}}\).

\subsection{General relativity limiting cases}\label{ssec:grlimit}
There are different limiting cases, in which the class of theories we consider reduces to general relativity. One of these cases is given by the teleparallel equivalent of general relativity (TEGR). This limit is obtained as a special case of the pure teleparallel action shown in section~\ref{ssec:puretele} by the choice
\begin{equation}
F(\mathcal{T}_1, \mathcal{T}_2, \mathcal{T}_3, X, Y, \phi, \mathbb{J}) = -\frac{1}{4}\mathcal{T}_1 - \frac{1}{2}\mathcal{T}_2 + \mathcal{T}_3\,,
\end{equation}
or equivalently by
\begin{equation}
G_{\text{Tele}}(\phi, X, T, T_{\text{ax}}, T_{\text{vec}}, I_2, \mathbb{J}) = -T\,.
\end{equation}
The same limit, up to a boundary term, can be achieved by starting from the Horndeski action discussed in section~\ref{ssec:horndeski} and choosing \(G_4 = 1, G_2 = G_3 = G_5 = 0\). In this case the only contribution to the action arises from the Lagrangian
\begin{equation}
\mathcal{L}_4 = -T + B\,,
\end{equation}
where the boundary term \(B\) does not contribute to the field equations. In both cases one finds the PPN parameters \(\beta = \gamma = 1\), as expected.

\subsection{Theories with $\beta = \gamma = 1$}\label{ssec:grmimic}
We finally determine the most general class of theories whose post-Newtonian limit agrees with the limit \(\beta = \gamma = 1\) obtained in general relativity. For this purpose, we start from the result~\eqref{eq:gammapar} for \(\gamma\). One immediately sees that this reduces to \(\gamma = 1\) if and only if
\begin{equation}
2\HH_{,1}\HH_{,4} + \HH_{,3}^2 = 0\,.
\end{equation}
We can classify the solutions of this equation into two branches:
\begin{enumerate}
\item
We first consider the case \(\HH_{,3} = 0\). In this case we must also demand \(\HH_{,4} = 0\) (since \(\HH_{,1}\) cancels in the expression~\eqref{eq:gammapar}), as well as \(\HH_{,5} \neq 0\) to avoid divergences due to strong coupling, as mentioned at the end of section~\ref{sec:PPN}. Inserting these conditions into our result~\eqref{eq:betapar} for \(\beta\) we find that also \(\beta = 1\). Theories of this type can be regarded as \emph{minimally coupled}, since neither the scalar field nor additional torsion terms contribute to the post-Newtonian limit.

We can also express this class of theories in the parametrization through the function \(F\). We find that the three conditions amount to
\begin{equation}\label{eq:mincoupgr}
\FF_{,Y} = 2\GG_{4,\phi}\,, \quad
2\FF_{,1} + \FF_{,2} + \FF_{,3} = 0\,, \quad
\FF_{,3} + \GG_4 \neq 0\,.
\end{equation}
In the parametrization through \(G_{\text{Tele}}\), the three conditions~\eqref{eq:mincoupgr} for the minimally coupled class of theories take the form
\begin{equation}
\GG_{\text{Tele},I_2} = 2\GG_{4,\phi}\,, \quad
\GG_{\text{Tele},T_{\text{vec}}} = 0\,, \quad
\GG_{\text{Tele},T} \neq \GG_4\,.
\end{equation}

\item
We can obtain another class of theories if we assume \(\HH_{,4} \neq 0\). This case may hence be regarded a \emph{non-minimally coupled} class of theories. In this case we may solve the condition \(\gamma = 1\) for, e.g., \(\HH_{,1}\) and find the Taylor coefficients
\begin{equation}
\HH_{,1} = -\frac{\HH_{,3}^2}{2\HH_{,4}}\,.
\end{equation}
Further, we must demand \(2\HH_{,4} - \HH_{,5} \neq 0\) to avoid divergences. Inserting this into the expression~\eqref{eq:betapar}, we find the simplified result
\begin{equation}
\beta - 1 =-\frac{\HH_{,4} \left(-3 \HH_{,2} \HH_{,3}^4+2 \HH_{,3}^4 \GG_{4,\phi}+2 \HH_{,4}^2 \HH_{,6}\right)}{4 \HH_{,3}^5 \left(2 \HH_{,4}-\HH_{,5}\right)} \,.
\end{equation}
Solving for \(\beta = 1\), we find the further conditions
\begin{equation}
\HH_{,6}=-\frac{\HH_{,3}^4 \left(2 \GG_{4,\phi}-3 \HH_{,2}\right)}{2 \HH_{,4}^2}\,,
\end{equation}
as well as \(\HH_{,3} \neq 0\).

In the parametrization through the function \(F\) this class is given by the assumptions
\begin{equation}
2\FF_{,1} + \FF_{,2} + \FF_{,3} \neq 0\,, \quad
2\FF_{,1} + \FF_{,2} - \GG_4 \neq 0\,, \quad
\FF_{,Y} \neq 2\GG_{4,\phi}\,,
\end{equation}
and solving for \(\gamma = 1\) yields the condition
\begin{equation}
\FF_{,X} = -\GG_{2,X} + 2\GG_{3,\phi} - \frac{1}{2}\frac{(\FF_{,Y} - 2\GG_{4,\phi})^2}{2\FF_{,1} + \FF_{,2} + \FF_{,3}}\,.\quad
\end{equation}
By using this condition in $\beta=1$, one gets an expression in terms of $\FF$, from which one can solve e.g.
\begin{multline}
\FF_{,\phi Y} = \frac{1}{2(2\GG_{4,\phi} - \FF_{,Y})(2\FF_{,1} + \FF_{,2} + \FF_{,3})}\Big[2(2\FF_{,1} + \FF_{,2} + \FF_{,3})^2(\GG_{2,\phi X} - 2\GG_{3,\phi\phi} + \FF_{,\phi X})\\
-4\GG_{4,\phi\phi}\FF_{,Y}(2\FF_{,1} + \FF_{,2} + \FF_{,3}) - \FF_{,Y}^2(6\FF_{,\phi 1} + 3\FF_{,\phi 2} + \FF_{,\phi 3}) - 4\GG_{4,\phi}^2(2\FF_{,Y} + 6\FF_{,\phi 1} + 3\FF_{,\phi 2} + \FF_{,\phi 3})\\
+ 8\GG_{4,\phi}^3 + 2\GG_{4,\phi}\left(\FF_{,Y}^2 + 4\GG_{4,\phi\phi}(2\FF_{,1} + \FF_{,2} + \FF_{,3}) + 2\FF_{,Y}(6\FF_{,\phi 1} + 3\FF_{,\phi 2} + \FF_{,\phi 3})\right)\Big]\,.
\end{multline}
Expressed through \(G_{\text{Tele}}\), the assumptions read
\begin{equation}
\GG_{\text{Tele},I_2} \neq 2\GG_{4,\phi}\,, \quad
\GG_{\text{Tele},T_{\text{vec}}} \neq 0\,, \quad
\GG_{\text{Tele},T} \neq \GG_4\,,
\end{equation}
and we can solve, e.g., for the Taylor coefficients for $\gamma=1$
\begin{equation}
\GG_{\text{Tele},X} = -\GG_{2,X} + 2\GG_{3,\phi} - \frac{(\GG_{\text{Tele},I_2} - 2\GG_{4,\phi})^2}{2\GG_{\text{Tele},T_{\text{vec}}}}
\end{equation}
while by replacing this expression in the condition $\beta=1$, one requires
\begin{eqnarray}
\GG_{\text{Tele},\phi T}&=&\frac{1}{2 \left(\GG_{\text{Tele},I_2}-2 \GG_{4,\phi}\right){}^2}\Big[2 \GG_{2,\phi X} \GG_{\text{Tele},T_{\text{vec}}}^2-4 \GG_{3,\phi\phi} \GG_{\text{Tele},T_{\text{vec}}}^2-4 \GG_{4,\phi}^2 \left(2 \GG_{\text{Tele},I_2}+\GG_{\text{Tele},\phi T_{\text{vec}}}\right)\nonumber\\
&&+2 \GG_{4,\phi} \Big(2 \GG_{\text{Tele},T_{\text{vec}}} \left(2 \GG_{4,\phi\phi}-\GG_{\text{Tele},\phi I_2}\right)+2 \GG_{\text{Tele},I_2} \GG_{\text{Tele},\phi T_{\text{vec}}}+\GG_{\text{Tele},I_2}^2\Big)+8 \GG_{4,\phi}^3\nonumber\\
&&+2 \GG_{\text{Tele},I_2} \Big(\GG_{\text{Tele},T_{\text{vec}}}( \GG_{\text{Tele},\phi I_2}-4 \GG_{4,\phi\phi})-\GG_{\text{Tele},I_2} \GG_{\text{Tele},\phi T_{\text{vec}}}\Big)+2 \GG_{\text{Tele},T_{\text{vec}}}^2 \GG_{\text{Tele},\phi X}\Big]
\end{eqnarray}
to have the same PPN parameters as in General Relativity.
\end{enumerate}

\section{Conclusion}
\label{sec:Conclusions}
We consider the post-Newtonian limit of a recently proposed covariant formulation of the teleparallel extension to the Horndeski class of gravity theories, called BDLS theory~\cite{Bahamonde:2019shr}. In order to apply the standard PPN formalism, we restrict ourselves to theories with a massless scalar field and in which terms of higher than second total derivative order are absent. We then calculate the PPN parameters for this restricted class of theories, and obtain a general formula for the PPN parameters in terms of the free functions determining the Lagrangian of a specific theory. Our findings show that the only PPN parameters which potentially deviate from their general relativity values are \(\gamma\) and \(\beta\), which means that all considered theories are fully conservative, i.e., they do not exhibit any preferred-frame or preferred-location effects or violation of total energy-momentum conservation. Further, we identify the class of theories whose parameters are identical to those of general relativity. We find a large class of theories, which are thus indistinguishable from general relativity by measuring the PPN parameters.

Our findings are in line with a number of previous results on the post-Newtonian limit of teleparallel gravity theories and generalize these previous results~\cite{Ualikhanova:2019ygl,Emtsova:2019qsl,Flathmann:2019khc}. By comparing our results with bounds on the PPN parameters obtained from solar system observations, we are able to constrain the class of teleparallel Horndeski theories, in addition to the constraints obtained from the speed of propagation of gravitational waves~\cite{Bahamonde:2019ipm} $->$ By comparing the results in this work with the observational values of the PPN parameters, we would be able to constraint model parameters for specific forms of the general BDLS theory. For instance, by using the Doppler tracking of the Cassini spacecraft about Saturn, it was found in Ref.\cite{Bertotti:2003rm} that $\gamma-1 = (2.1 \pm 2.3) \times 10^{-5}$, while in Ref.\cite{Fienga:2011qh} the perihelion advance of Mars can be used to obtain $\beta - 1 = (0.4 \pm 2.4)\times10^{-4}$. These constraints may be crucial for selecting viable models within the framework of models that are possible in general BDLS theory. Most interestingly, we find a large class of theories beyond general relativity, which pass all solar system tests of the PPN parameters.

These results further consolidate the BDLS class of teleparallel gravity theories as an interesting alternative to curvature-based modified gravity theories, and motivate further studies of their properties, as well as similar theories whose construction is based on related principles~\cite{Hohmann:2019gmt}. Our calculation of the parametrized post-Newtonian limit can be seen as a first step towards a more general post-Newtonian approximation to higher perturbation orders, which allows to derive the gravitational waves emitted from orbiting compact objects~\cite{Blanchet:2013haa}. Complementary sets of observables may be derived by using the theory of cosmological perturbations~\cite{Kodama:1985bj,Mukhanov:1990me,Malik:2008im} to determine, e.g., the parameters of inflation; using universal relations between the observable properties of neutron stars~\cite{Yagi:2013bca,Yagi:2013awa}; studying the propagation of light and the existence of photon regions around compact objects~\cite{Cunha:2018acu}. From the theoretical side, one may employ the Hamiltonian formalism to study the appearance of additional degrees of freedom, further generalizing earlier results~\cite{Ferraro:2016wht,Ferraro:2018tpu,Blixt:2018znp}. Combining these different directions of research will allow us to further constrain the class of viable teleparallel gravity models.

\section*{Acknowledgements}
The authors would like to acknowledge networking support by the COST Action GWverse CA16104 and QG-MM CA18108. This article is based upon work from CANTATA COST (European Cooperation in Science and Technology) action CA15117, EU Framework Programme Horizon 2020. S.B. is supported by Mobilitas Pluss No. MOBJD423 by the Estonian government. M.H. gratefully acknowledges the full financial support by the Estonian Research Council through the Personal Research Funding projects PRG356 and PSG489. S.B. and M.H. gratefully acknowledge the full financial support by the European Regional Development Fund through the Center of Excellence TK133 ``The Dark Side of the Universe''. JLS would like to acknowledge funding support Cosmology@MALTA which is supported by the University of Malta.
\appendix
\section{Derivation of the field equations}\label{appendix1}
\subsection{Useful identities}\label{sec:basic}
To be able to find the field equations for our model, we will use some the following identities~\cite{Bahamonde:2017wwk}
\begin{eqnarray}
\frac{\partial e^B{}_{ \nu}}{\partial e^A{}_{\mu}} &=&
\delta^B_A \delta^\mu_\nu\,, \qquad \frac{\partial E_B{}^{ \nu}}{\partial e^{A}{}_{\mu}} =
- E_B{}^{\mu} E_A{}^{ \nu}\,,
\label{id1}\\
\frac{\partial e}{\partial e^{A}{}_{\mu}}&=&e E_A{}^{\mu} \,, \qquad
\frac{\partial g^{\alpha\beta}}{\partial e^A{}_{\mu}} =
-g^{\mu\beta} E_A{}^{\alpha}  -
g^{\mu\alpha} E_A{}^{\beta}\,, \quad \frac{\partial g_{\alpha\beta}}{\partial e^A{}_{\mu}} =
\eta_{AB}\delta^\mu_\alpha e^B{}_\beta+\eta_{AB}\delta^\mu_\beta e^B{}_\alpha\,,\label{id2}
\end{eqnarray}
In addition, for variations of the torsion tensor we can find
\begin{align}
\label{id3}
\frac{\partial T^B{}_{ \rho\sigma}}{\partial e^A{}_{\mu}} &= \omega^B{}_{A \rho}\delta^\mu_\sigma -\omega^B{}_{A \sigma}\delta^\mu_\rho\,,  \qquad
\frac{\partial T^B{}_{\rho\sigma}}{\partial e^A{}_{\mu,\nu}} = \delta_A^B(\delta_\rho^\nu \delta_\sigma^\mu - \delta_\sigma^\nu \delta_\rho^\mu)\,.
\end{align}
For the vectorial part of torsion, one has
\begin{align}
C_{IA}{}^{\mu}\equiv \frac{\partial v_I}{\partial e^A{}_{\mu}}&={E_I{}^{ \mu} \omega^\rho{}_{ A \rho}	- \omega^\mu{}_{A I} -	T^\mu{}_{ AI}     	-	v_A	E_I{}^{ \mu} }\,,\label{vector}\\
 \frac{\partial v_I}{\partial e^A{}_{ \mu,\nu}} &=
{E_A{}^{ \nu } E_I{}^{ \mu}  - E_A{}^{ \mu } E_I{}^{ \nu} }\,.\label{vector2}
\end{align}
On the other hand, for the axial part, one can compute the following identities
\begin{eqnarray}
\frac{\partial a_I}{\partial e^A{}_{ \mu}}
&=&{
	-\frac{1}{3}\epsilon_{IB}{}^{ CD}
	E_C{}^{ \mu}  T^{B}{}_{ AD}
	+
	\frac{1}{3}\epsilon_{IB}{}^{CD}
	E_D{}^{ \mu} \omega^B{}_{ AC}}\,,\label{axial1}\\
\frac{\partial a_I}{\partial e^A{}_{ \mu,\nu}}
&=&{\frac{1}{3}\epsilon_{IA}{}^{ CD}   E_C{}^{ \nu}E_D{}^{ \mu} }\,.\label{axial}
\end{eqnarray}
The important derivatives for the tensorial part are more involved and were not presented in~\cite{Bahamonde:2017wwk} so, we are going to find them here. The variations of the tensorial part with respect to the tetrads gives us
\begin{eqnarray}
\frac{\partial t_{IJK}}{\partial e^A{}_{ \mu}}=\frac{1}{2}\frac{\partial }{\partial e^A{}_{ \mu}}\Big[T_{IJK}+T_{JIK}\Big]+\frac{1}{6}\frac{\partial }{\partial e^A{}_{ \mu}}\Big[\eta_{KI}v_J-\eta_{KJ}v_I-2\eta_{IJ}v_K\Big]\,.\label{ola}
\end{eqnarray}
The first term in the first bracket can be written as
\begin{eqnarray}
\frac{\partial }{\partial e^A{}_{ \mu}}T_{IJK}=\frac{\partial }{\partial e^A{}_{ \mu}}(T^{B}{}_{\rho\sigma}E_J{}^{\rho}E_K{}^{\sigma})=\omega_{IAJ}E_K{}^{\mu}-\omega_{IAK}E_J{}^{\mu}-T_{IJA}E_K{}^{\mu}-T_{IAK}E_J{}^\mu\,,\label{Tis}
\end{eqnarray}
where we have used Eqs.~\eqref{id1}--\eqref{id3}. The second term in~\eqref{ola} can be written as
\begin{eqnarray}
\frac{1}{6}\frac{\partial }{\partial e^A{}_{ \mu}}\Big[\eta_{KI}v_J-\eta_{KJ}v_I-2\eta_{IJ}v_K\Big]&=&\frac{1}{6}\Big[\eta_{KI}C_{JA}{}^{\mu}-\eta_{KJ}C_{IA}{}^{\mu}-2\eta_{IJ}C_{KA}{}^{\mu}+v_J D_{KIA}{}^{\mu}-v_I D_{KJA}{}^{\mu}\nonumber\\
&&-2v_K D_{IJA}{}^{\mu}\Big]\,,\label{ola2}
\end{eqnarray}
where $C_{IA}{}^{\mu}$ was defined in~\eqref{vector} and $D_{KIA}{}^{\mu}$ is given by
\begin{eqnarray}
D_{KIA}{}^{\mu}\equiv \frac{\partial \eta_{KI}}{\partial e^A{}_\mu}=\delta^B_I \eta_{AB}E^\mu{}_K+\delta^B_K \eta_{AB}E^\mu{}_I-\eta_{AI}E^\mu{}_K-\eta_{KA}E^\mu{}_I\,.
\end{eqnarray}
Thus, by replacing~\eqref{Tis} and \eqref{ola2} into \eqref{ola} one gets
\begin{eqnarray}
\frac{\partial t_{IJK}}{\partial e^A{}_{ \mu}}&=&\frac{1}{2}\Big[\omega_{IAJ}E_K{}^{\mu}-\omega_{IAK}E_J{}^{\mu}-T_{IJA}E_K{}^{\mu}-T_{IAK}E_J{}^\mu+\omega_{JAI}E_K{}^{\mu}-\omega_{JAK}E_I{}^{\mu}-T_{JIA}E_K{}^{\mu}-T_{JAK}E_I{}^\mu\Big]\nonumber\\
&&+\frac{1}{6}\Big[\eta_{KI}C_{JA}{}^{\mu}-\eta_{KJ}C_{IA}{}^{\mu}-2\eta_{IJ}C_{KA}{}^{\mu}+v_JD_{KIA}{}^{\mu}-v_ID_{KJA}{}^{\mu}-2v_K D_{IJA}{}^{\mu}\Big]\equiv H_{IJKA}{}^{\mu}\,.\label{tensorial0}
\end{eqnarray}
The last important identity needed is
\begin{eqnarray}
\frac{\partial t_{IJK}}{\partial e^A{}_{ \mu,\nu}}&=&\frac{1}{2}\frac{\partial }{\partial e^A{}_{ \mu,\nu}}\Big[T_{IJK}+T_{JIK}\Big]+\frac{1}{6}\frac{\partial }{\partial e^A{}_{ \mu,\nu}}\Big[\eta_{KI}v_J-\eta_{KJ}v_I-2\eta_{IJ}v_K\Big]\,.\label{tensorial1}
\end{eqnarray}
The first term in bracket can be  easily find,
\begin{eqnarray}
\frac{\partial }{\partial e^A{}_{ \mu,\nu}}T_{IJK}=\frac{\partial }{\partial e^A{}_{ \mu,\nu}}(T^{B}{}_{\rho\sigma}\eta_{BI}E_J{}^{\rho}E_K{}^{\sigma})=\eta_{BI}E_J{}^{\rho}E_K{}^{\sigma}\frac{\partial }{\partial e^A{}_{ \mu,\nu}}(T^{B}{}_{\rho\sigma})=\eta_{AI}(E_J{}^{\nu}E_K{}^{\mu}-E_J{}^{\mu}E_K{}^{\nu})\,,\label{ola3}
\end{eqnarray}
where we have used \eqref{id3}. The second term in~\eqref{tensorial1} is
\begin{eqnarray}
\frac{\partial }{\partial e^A{}_{ \mu,\nu}}(\eta_{KI}v_J)=\eta_{KI}\frac{\partial }{\partial e^A{}_{ \mu,\nu}}v_J=\eta_{KI}(E_A{}^{ \nu } E_J{}^{ \mu}  - E_A{}^{ \mu } E_J{}^{ \nu} )\,.
\end{eqnarray}
If one replaces the above identity and also \eqref{ola3} into \eqref{tensorial1}, one finally gets
\begin{eqnarray}
\frac{\partial t_{IJK}}{\partial e^A{}_{ \mu,\nu}}&=&\frac{1}{2}\Big[\eta_{AI}(E_J{}^{\nu}E_K{}^{\mu}-E_J{}^{\mu}E_K{}^{\nu})+\eta_{AJ}(E_I{}^{\nu}E_K{}^{\mu}-E_I{}^{\mu}E_K{}^{\nu})\Big]+\frac{1}{6}\Big[\eta_{KI}(E_A{}^{ \nu } E_J{}^{ \mu}  - E_A{}^{ \mu } E_J{}^{ \nu})\nonumber\\
&&-\eta_{KJ}(E_A{}^{ \nu } E_I{}^{ \mu}  - E_A{}^{ \mu } E_I{}^{ \nu})-2\eta_{IJ}(E_A{}^{ \nu } E_K{}^{ \mu}  - E_A{}^{ \mu } E_K{}^{ \nu})\Big]\equiv L_{IJKA}{}^{\mu\nu}\,.\label{tensorial}
\end{eqnarray}
\subsection{Variations with respect to the tetrads}\label{sec:tetrad}
\subsubsection{Variations of $\mathcal{L}_{\rm Tele}$}
The term $e \delta_{e} \mathcal{L}_{\rm Tele}$ can be expanded as
\begin{eqnarray}
e \delta_{e} \mathcal{L}_{\rm Tele}=e \delta_e G_{\rm Tele}&=&e \Big(G_{\rm Tele,X} \delta_e X+G_{\rm Tele,T} \delta_e T+G_{\rm Tele,T_{\rm ax}} \delta_e T_{\rm ax}+G_{\rm Tele,T_{\rm vec}} \delta_e T_{\rm vec}+G_{\rm Tele,I_2} \delta_e I_2+G_{\rm Tele,J_1} \delta_e J_1\nonumber\\
&&+G_{\rm Tele,J_3} \delta_e J_3+G_{\rm Tele,J_5} \delta_e J_5+G_{\rm Tele,J_6} \delta_e J_6+G_{\rm Tele,J_8} \delta_e J_8+G_{\rm Tele,J_{10}} \delta_e J_{10}\Big)\,.\label{Eqa}
\end{eqnarray}
Since all these scalars are constructed from $a_I, v_I$ and $t_{IJK}$ and they contain up to first derivatives in the tetrads, one can notice that, the only important terms appearing in the field equations are
\begin{eqnarray}
C^I\delta v_I&=&\Big[C^I\frac{\partial v_I }{\partial e^A{}_{ \mu}}-\partial_\nu\Big(C^I\frac{\partial v_I }{\partial e^{A}{}_{  \mu,\nu}}\Big)\Big]\delta e^A{}_\mu\,,\label{var1}\\
C^I\delta a_I&=&\Big[C^I\frac{\partial a_I }{\partial e^A{}_{ \mu}}-\partial_\nu\Big(C^I\frac{\partial a_I }{\partial e^{A}{}_{  \mu,\nu}}\Big)\Big]\delta e^A{}_\mu\,,\label{var2}\\
C^{IJK}\delta t_{IJK}&=&\Big[C^{IJK}\frac{\partial t_{IJK} }{\partial e^A{}_{ \mu}}-\partial_\nu\Big(C^{IJK}\frac{\partial t_{IJK} }{\partial e^{A}{}_{  \mu,\nu}}\Big)\Big]\delta e^A{}_\mu\,,\label{var3}
\end{eqnarray}
where $C^I$ and $C^{IJK}$ are any arbitrary vector and tensor respectively. Note that the above derivatives for $v_I,a_I$ and $t_{IJK}$  are given in~\eqref{vector}, \eqref{vector2}, \eqref{axial1}, \eqref{axial}, \eqref{tensorial0} and \eqref{tensorial}, respectively. The first two above variations are
\begin{eqnarray}
eG_{\rm Tele,X} \delta_e X&=&-\frac{1}{2}eG_{\rm Tele,X}\phi_{;\alpha}\phi_{;\beta} \delta_e  g^{\alpha\beta}=eG_{\rm Tele,X}\phi_{;\alpha}\phi^{;\mu} E_A{}^{\alpha}\delta e^{A}{}_\mu\,,\label{X}\\
eG_{\rm Tele,T} \delta_e T&=&-4e\Big[(\partial_{\lambda}G_{\rm Tele,T})S_{A}\,^{\lambda\mu}+e^{-1}\partial_{\lambda}(e S_{A}\,^{\lambda\mu})G_{\rm Tele,T}-G_{\rm Tele,T}T^{\sigma}\,_{\lambda A}S_{\sigma}\,^{\mu\lambda}\nonumber\\
&&+G_{\rm Tele,T}\omega^{B}{}_{A\nu}S_{B}{}^{\nu\mu}\Big]\delta e^{A}_{\mu}\,,\label{T}
\end{eqnarray}
where we have used the result found in~\cite{Krssak:2015oua,Bahamonde:2015zma} to compute the second term $eG_{\rm Tele,T} \delta_e T$ and Eq.~\eqref{id2} to compute the first one. All the remaining variations are given by
\begin{eqnarray}
eG_{\rm Tele,T_{\rm ax}} \delta_e T_{\rm ax}&=&2eG_{\rm Tele,T_{\rm ax}} a^I\delta_e a_I\,,\label{Taxvar}\\
eG_{\rm Tele,T_{\rm vec}} \delta_e T_{\rm vec}&=&2eG_{\rm Tele,T_{\rm vec}} v^I\delta_e v_I\,,\\
eG_{\rm Tele,I_2} \delta_e I_2&=&eG_{\rm Tele,I_2}\phi^{;I} \delta_e v_{I}-eG_{\rm Tele,I_2}v^\mu \phi_{;A}  \delta_e e^{A}{}_\mu\,,\\
eG_{\rm Tele,J_1} \delta_e J_1&=&2eG_{\rm Tele,J_1}\phi^{;I}\phi^{;J}a_{J} \delta_e a_{I}-2eG_{\rm Tele,J_1}a_{J}a^{\mu}\phi^{;J}\phi_{;A}  \delta_e e^{A}{}_\mu \,,\\
eG_{\rm Tele,J_3} \delta_e J_3&=&eG_{\rm Tele,J_3}\phi^{;K}\phi^{;J}t^{I}{}_ {KJ}\delta_e  v_{I}+eG_{\rm Tele,J_3}\phi^{;J}\phi^{;K}v^{I}\delta_e t_{IJK} \nonumber\\
&&+eG_{\rm Tele,J_3}v_I t_{K}{}^{\mu I}\phi^{;K}\phi_{;A} \delta_e e^{A}{}_\mu\,, \\
eG_{\rm Tele,J_5} \delta_e J_5&=&2eG_{\rm Tele,J_5}\phi^{;L}\phi^{;J}t^{I}{}_{L}{}^{K}\delta_e t_{IJK}-2eG_{\rm Tele,J_5} t^{I\mu K}t_{IJK}\phi^{;J}\phi_{;A}\delta_e e^{A}{}_\mu\,,\\
eG_{\rm Tele,J_6} \delta_e J_6&=&2eG_{\rm Tele,J_6}\phi^{;J}\phi^{;K}\phi^{;L}\phi^{;M}t^{I}{}_{LM}\delta_e t_{IJK}+2eG_{\rm Tele,J_6}t_{ILK}t^\mu{}_M{}^I\phi^{;K}\phi^{;L}\phi^{;M}\phi_{;A}  \delta_e e^{A}{}_\mu\,,\\
eG_{\rm Tele,J_8} \delta_e J_8&=&2eG_{\rm Tele,J_8}\phi^{;L}\phi^{;K}t^{IJ}{}_{L}\delta_e t_{IJK}-2eG_{\rm Tele,J_8}t_{IJK}t^{IJ}{}^{\mu}\phi^{;K}\phi_{;A}  \delta_e e^{A}{}_\mu\,,\label{J8var}\\
G_{\rm Tele,J_{10}} \delta_e J_{10}&=&eG_{\rm Tele,J_{10}}\epsilon_{A}{}^{I}{}_{CD}\phi^{;A}\phi^{;J}t_{J}{}^{CD}\delta_e a_I+eG_{\rm Tele,J_{10}}\epsilon_{AB}{}^{JK}A^B \phi^{;A}\phi^{;I}\delta_e t_{IJK}\nonumber\\
&&-eG_{\rm Tele,J_{10}}a^J \phi^{;I}\phi_{;A}\Big(\epsilon_{\mu JCD}t_I{}^{CD}+\epsilon_{IJCD}t^{\mu CD}\Big) \delta_e e^{A}{}_\mu\,.
\end{eqnarray}
Here, we have used the identity $t^{\mu\nu\alpha}=-t^{\alpha\mu\nu}-t^{\alpha\nu\mu}$ and also Eq.~\eqref{id1} to replace $\delta_e E_B{}^\sigma=-E_{B}{}^\mu E_A{}^\sigma \delta_e e^{A}{}_\mu$. Thus, by replacing~\eqref{Taxvar}--\eqref{J8var} into \eqref{Eqa} one can rewrite the variations as
\begin{eqnarray}
e \delta_{e} \mathcal{L}_{\rm Tele}&=&eG_{\rm Tele,T} \delta_e T+e\Big[G_{\rm Tele,X}\phi^{;\mu}-G_{\rm Tele,I_2}v^\mu   -2G_{\rm Tele,J_1}a^{\mu}a_{J}\phi^{;J} +G_{\rm Tele,J_3}v_I t_{K}{}^{\mu I}\phi^{;K}-2G_{\rm Tele,J_5} t^{I\mu K}t_{IJK}\phi^{;J}\nonumber\\
&&+2G_{\rm Tele,J_6}t_{ILK}t^\mu{}_M{}^I\phi^{;K}\phi^{;L}\phi^{;M}-2G_{\rm Tele,J_8}t_{IJK}t^{IJ}{}^{\mu}\phi^{;K}-G_{\rm Tele,J_{10}}a^J \phi^{;I}\Big(\epsilon_{\mu JCD}t_I{}^{CD}\nonumber\\
&&+\epsilon_{IJCD}t^{\mu CD}\Big) \Big] \phi_{;A}\delta e^A{}_\mu+eM^I\delta_{e}a_I+eN^I\delta_{e}v_I+eO^{IJK}\delta_{e}t_{IJK}\,,
\end{eqnarray}
where we have defined the following quantities
\begin{eqnarray}
M^{I}&=&2G_{\rm Tele,T_{\rm ax}} a^I+2G_{\rm Tele,J_1}\phi^{;I}\phi^{;J}a_{J}+G_{\rm Tele,J_{10}}\epsilon_{A}{}^{I}{}_{CD}\phi^{;A}\phi^{;J}t_{J}{}^{CD}\,,\label{M}\\
N^I&=&2G_{\rm Tele,T_{\rm vec}} v^I+G_{\rm Tele,I_2}\phi^{;I}+2G_{\rm Tele,J_2}\phi^{;I}\phi^{;J}v_{J}+G_{\rm Tele,J_3}\phi^{;K}\phi^{;J}t^{I}{}_ {KJ}\,,\label{N}\\
O^{IJK}&=&G_{\rm Tele,J_3}\phi^{;J}\phi^{;K}v^{I}+2G_{\rm Tele,J_5}\phi^{;L}\phi^{;J}t^{I}{}_{L}{}^{K}+2G_{\rm Tele,J_6}\phi^{;J}\phi^{;K}\phi^{;L}\phi^{;M}t^{I}{}_{LM}+2G_{\rm Tele,J_8}\phi^{;L}\phi^{;K}t^{IJ}{}_{L}\nonumber\\
&&+G_{\rm Tele,J_{10}}\epsilon_{AB}{}^{JK} \phi^{;A}\phi^{;B}\phi^{;I}\,.\label{O}
\end{eqnarray}
Now by using Eqs.~\eqref{var1}-\eqref{var3}, one gets
\begin{eqnarray}
e \delta_{e} \mathcal{L}_{\rm Tele}&=&eG_{\rm Tele,T} \delta_e T+e\Big[G_{\rm Tele,X}\phi^{;\mu}-G_{\rm Tele,I_2}v^\mu   -2G_{\rm Tele,J_1}a^{\mu}a_{J}\phi^{;J} +G_{\rm Tele,J_3}v_I t_{K}{}^{\mu I}\phi^{;K}\nonumber\\
&&-2G_{\rm Tele,J_5} t^{I\mu K}t_{IJK}\phi^{;J}+2G_{\rm Tele,J_6}t_{ILK}t^\mu{}_M{}^I\phi^{;K}\phi^{;L}\phi^{;M}-2G_{\rm Tele,J_8}t_{IJK}t^{IJ}{}^{\mu}\phi^{;K}\nonumber\\
&& -G_{\rm Tele,J_{10}}a^J \phi^{;I}\Big(\epsilon_{\mu JCD}t_I{}^{CD}+\epsilon_{IJCD}t^{\mu CD}\Big)\Big] \phi_{;A}\delta e^A{}_\mu+\Big[eM^I\frac{\partial a_I }{\partial e^A{}_{ \mu}}-\partial_\nu\Big(eM^I\frac{\partial a_I }{\partial e^{A}{}_{  \mu,\nu}}\Big)\Big]\delta e^A{}_\mu\nonumber\\
&&+\Big[eN^I\frac{\partial v_I }{\partial e^A{}_{ \mu}}-\partial_\nu\Big(eN^I\frac{\partial v_I }{\partial e^{A}{}_{  \mu,\nu}}\Big)\Big]\delta e^A{}_\mu+\Big[eO^{IJK}\frac{\partial t_{IJK} }{\partial e^A{}_{ \mu}}-\partial_\nu\Big(eO^{IJK}\frac{\partial t_{IJK} }{\partial e^{A}{}_{  \mu,\nu}}\Big)\Big]\delta e^A{}_\mu\,,
\end{eqnarray}
which can be written explicitly using the identities~\eqref{vector}--\eqref{tensorial}, yielding
\begin{eqnarray}
e \delta_{e} \mathcal{L}_{\rm Tele}&=&\Big[-4e\Big\{(\partial_{\lambda}G_{\rm Tele,T})S_{A}\,^{\lambda\mu}+e^{-1}\partial_{\lambda}(e S_{A}\,^{\lambda\mu})G_{\rm Tele,T}-G_{\rm Tele,T}T^{\sigma}\,_{\lambda A}S_{\sigma}\,^{\mu\lambda}+G_{\rm Tele,T}\omega^{B}{}_{A\nu}S_{B}{}^{\nu\mu}\Big\}\nonumber\\
&&+e\,\phi_{;A}\Big\{G_{\rm Tele,X}\phi^{;\mu}-G_{\rm Tele,I_2}v^\mu   -2G_{\rm Tele,J_1}a^{\mu}a_{J}\phi^{;J} +G_{\rm Tele,J_3}v_I t_{K}{}^{\mu I}\phi^{;K}-2G_{\rm Tele,J_5} t^{I\mu K}t_{IJK}\phi^{;J}\nonumber\\
&&+2G_{\rm Tele,J_6}t_{ILK}t^\mu{}_M{}^I\phi^{;K}\phi^{;L}\phi^{;M}-2G_{\rm Tele,J_8}t_{IJK}t^{IJ}{}^{\mu}\phi^{;K} -G_{\rm Tele,J_{10}}a^J \phi^{;I}\Big(\epsilon_{\mu JCD}t_I{}^{CD}+\epsilon_{IJCD}t^{\mu CD}\Big) \Big\} \nonumber\\
&&+\frac{1}{3}\Big\{eM^I(
\epsilon_{IB}{}^{ CD}
E_D{}^{ \mu} \omega^B{}_{ AC}-\epsilon_{IB}{}^{ CD}
E_C{}^{ \mu}  T^{B}{}_{ AD} )-\partial_\nu\Big(eM^I\epsilon_{IA}{}^{ CD}   E_C{}^{ \nu}E_D{}^{ \mu}\Big)\Big\}\nonumber\\
&&+eN^I(E_I{}^{ \mu} \omega^\rho{}_{ A \rho}	- \omega^\mu{}_{AI} -	T^\mu{}_{AI}     	-	v_A	E_I{}^{ \mu})-\partial_\nu\Big(eN^I(E_A{}^{ \nu } E_I{}^{ \mu}  - E_A{}^{ \mu } E_I{}^{ \nu})\Big)\nonumber\\
&&+eO^{IJK}H_{IJKA}{}^{\mu}-\partial_\nu\Big(eO^{IJK}L_{IJKA}{}^{\mu\nu}\Big)\Big]\delta e^A{}_\mu\,,\label{Finalvariation}
\end{eqnarray}
where we have used \eqref{T}, and $H_{IJKA}{}^{\mu}$ and $L_{IJKA}{}^{\mu\nu}$ were defined in Eqs.~\eqref{tensorial0} and \eqref{tensorial}, respectively. The above equation is the contribution coming from the Teleparallel term $\mathcal{L}_{\rm Tele}$.

\subsubsection{Variations of $\mathcal{L}_{i}$  ($i=2,..,5$)}
The Lagrangians $\mathcal{L}_{i}$ ($i=2,..,5$) are exactly the same as the standard Horndeski gravity theory, therefore, it is not necessary to compute the variations of the field equations again. The variations for these terms were computed in~\cite{Capozziello:2018gms} but in terms of the metric. It is easy to modify them by using the identity \eqref{id1}--\eqref{id3}, i.e., by replacing $\delta g^{\alpha\beta}=-(g^{\mu\beta}E_A{}^{\alpha}+g^{\mu\alpha}E_A{}^{\beta})\delta e_A{}^{\mu}$. In other words, Eqs.~(11.a)--(13.d) in \cite{Capozziello:2018gms} are explicitly the field equations for these terms. To convert these terms into our tetrad notation, we need to take $\sum_{i=2}^{5}\mathcal{G}_{\mu\nu}^{(i)}$ in \cite{Capozziello:2018gms} and convert it to $-2\sum_{i=2}^{5}\mathcal{G}^{(i)}{}_A{}^{\mu}$. The factor $-2$ comes from the fact that the variations with respect to the tetrad give rise two minus terms when one is changing into metric variations ($\delta g^{\alpha\beta}=-(g^{\mu\beta}E_A{}^{\alpha}+g^{\mu\alpha}E_A{}^{\beta})\delta e_A{}^{\mu}$). Thus, the variations that appears in \eqref{deltaS} will be equal to
\begin{eqnarray}
e\sum_{i=2}^{5}\mathcal{L}_iE_A{}^{\mu}\delta e^A{}_\mu+e \delta_{e} \sum_{i=2}^{5}\mathcal{L}_{i}=-2e\sum_{i=2}^{5}\mathcal{G}^{(i)}{}_A{}^{\mu}\delta e^{A}{}_\mu=-2eE_A{}^{\nu}g^{\mu\alpha}\sum_{i=2}^{5}\mathcal{G}^{(i)}{}_{\alpha\nu}\delta e^{A}{}_\mu\,,\label{GG}
\end{eqnarray}
where $\mathcal{G}^{(i)}{}_{\alpha\nu}$ were explicitly found in Eqs.~(13a)--(13d) in \cite{Capozziello:2018gms}.

\subsection{Variations with respect to the scalar field}\label{sec:varphi}
For the Teleparallel Horndeski Lagrangian, one can expand its variations with respect to the scalar field as
\begin{eqnarray}
\delta_{\phi}(e\mathcal{L}_{\rm Tele})&=&eG_{\rm Tele,\phi}\delta_\phi\phi+eG_{\rm Tele,X}\delta_\phi X+eG_{\rm Tele,I_2}\delta_\phi I_2+eG_{\rm Tele,J_1}\delta_\phi J_1+eG_{\rm Tele,J_3}\delta_\phi J_3\nonumber\\
&&+eG_{\rm Tele,J_5}\delta_\phi J_5+eG_{\rm Tele,J_8}\delta_\phi J_8+eG_{\rm Tele,J_6}\delta_\phi J_6+eG_{\rm Tele,J_{10}}\delta_\phi J_{10}\,.\label{varphi}
\end{eqnarray}
All the other invariants do not depend on the scalar field, hence, their variations are identically zero. The first five terms can be straightforwardly computed, yielding
\begin{eqnarray}
eG_{\rm Tele,\phi}\delta_\phi\phi&=&eG_{\rm Tele,\phi}\delta \phi\,,\\
eG_{\rm Tele,X}\delta_\phi X&=&-\frac{1}{2}eG_{\rm Tele,X}\delta_\phi \Big[g^{\mu\nu}(\partial_\mu\phi)(\partial_\nu\phi)\Big]=\partial_\mu\Big[eG_{\rm Tele,X}g^{\mu\nu}(\partial_\nu\phi)\Big]\delta \phi\,,\\
eG_{\rm Tele,I_2}\delta_\phi I_2&=&eG_{\rm Tele,I_2}\delta_\phi (v^\mu \partial_\mu \phi)=-\partial_\mu(eG_{\rm Tele,I_2} v^\mu)\delta\phi\,,\\
eG_{\rm Tele,J_1}\delta_\phi J_1&=&eG_{\rm Tele,J_1}\delta_\phi (a^\mu a^\nu \partial_\mu \partial_\nu \phi)=-2\partial_\mu (eG_{\rm Tele,J_1}a^\mu a^\nu\partial_\nu \phi)\delta \phi\,,\label{varphi2}
\end{eqnarray}
where we have integrated by parts. The next three terms in~\eqref{varphi} become
\begin{eqnarray}
eG_{\rm Tele,J_3}\delta_\phi J_3&=&eG_{\rm Tele,J_3}\delta_\phi\Big[ v_{\alpha}t^{\alpha\mu\nu}(\partial_\mu\phi)(\partial_\nu\phi)\Big]=\partial_\mu\Big[eG_{\rm Tele,J_3}v_\alpha t^{\mu\nu\alpha}(\partial_\nu\phi)\Big]\delta\phi\,,\\
eG_{\rm Tele,J_5}\delta_\phi J_5&=&eG_{\rm Tele,J_5}\delta_\phi \Big[t^{\mu\nu\alpha}t_{\mu}{}^{\lambda}{}_\alpha (\partial_\nu\phi)(\partial_\lambda\phi) \Big]=-2\partial_\mu\Big[eG_{\rm Tele,J_5}t^{\beta\nu\alpha}t_{\beta}{}^{\mu}{}_\alpha(\partial_\nu\phi)\Big]\delta\phi\,,\\
eG_{\rm Tele,J_8}\delta_\phi J_8&=&eG_{\rm Tele,J_8}\delta_\phi \Big[t^{\mu\nu\alpha}t_{\mu\nu}{}^{\beta} (\partial_\alpha\phi)(\partial_\beta\phi) \Big]=-2\partial_\mu\Big[eG_{\rm Tele,J_8}t^{\alpha\nu\mu}t_{\alpha\nu}{}^{\beta}(\partial_\beta\phi)\Big]\delta\phi\,,\label{varphi3}\\
eG_{\rm Tele,J_{10}}\delta_\phi J_{10}&=&eG_{\rm Tele,J_{10}}\delta_\phi \Big[\epsilon^{\mu}{}_{\nu\rho\sigma}a^\nu t^{\alpha\rho\sigma}\phi_{;\mu}\phi_{;\alpha} \Big]=-\partial_{\mu}\Big[eG_{\rm Tele,J_{10}} a^\nu (\partial_\alpha \phi) (\epsilon^{\mu}{}_{\nu\rho\sigma}t^{\alpha\rho\sigma}+\epsilon^{\alpha}{}_{\nu\rho\sigma}t^{\mu\rho\sigma})\Big]\delta \phi\label{varphi3b}\,,
\end{eqnarray}
where we have used the identity $t^{\mu\nu\alpha}=-t^{\alpha\mu\nu}-t^{\alpha\nu\mu}$, the symmetry property of the tensorial part of torsion $t^{\mu\nu\alpha}=t^{\nu\mu\alpha}$, and we have integrated by parts. The last term in \eqref{varphi} is more involved but it also can be directly computed, giving
\begin{eqnarray}
eG_{\rm Tele,J_6}\delta_\phi J_6&=&eG_{\rm Tele,J_6}\delta_\phi \Big[t^{\mu\nu\alpha}t_{\mu}{}^{\beta\sigma}(\partial_\alpha\phi)(\partial_\beta\phi) (\partial_\nu\phi)(\partial_\sigma\phi) \Big]\\
&=&2\partial_\mu\Big[eG_{\rm Tele,J_6}t^{\nu\alpha\beta}t^{\mu\sigma}{}_\nu(\partial_\alpha\phi)(\partial_\beta\phi)(\partial_\sigma\phi)\Big]\delta\phi\,,\label{varphi4}
\end{eqnarray}
where again we have used the identities $t^{\mu\nu\alpha}=-t^{\alpha\mu\nu}-t^{\alpha\nu\mu}$ and $t^{\mu\nu\alpha}=t^{\nu\mu\alpha}$ several times and we have ignored boundary terms. Thus, by replacing all the variations given by \eqref{varphi2}--\eqref{varphi4} into \eqref{varphi}, we find the final expression for the variation of the Teleparallel Lagrangian with respect to the scalar field which reads as follows
\begin{eqnarray}
\delta_{\phi}(e\mathcal{L}_{\rm Tele})&=&-e\Big[\mathring{\nabla}^\mu (J_{\mu\rm -Tele})-P_{\phi-\rm Tele}\Big]\delta\phi\,,\label{phitele}
\end{eqnarray}
where we have defined
\begin{eqnarray}
J_{\mu\rm -Tele}&=&-G_{\rm Tele,X}(\mathring{\nabla}_\mu\phi)+G_{\rm Tele,I_2} v_\mu+2G_{\rm Tele,J_1}a_\mu a^\nu\mathring{\nabla}_\nu \phi-G_{\rm Tele,J_3}v_\alpha t_{\mu}{}^{\nu\alpha}(\mathring{\nabla}_\nu\phi)\nonumber\\
&&-2G_{\rm Tele,J_5}t^{\beta\nu\alpha}t_{\beta\mu\alpha}(\mathring{\nabla}_\nu\phi)+2G_{\rm Tele,J_8}t^{\alpha\nu}{}_{\mu}t_{\alpha\nu}{}^{\beta}(\mathring{\nabla}_\beta\phi)-2G_{\rm Tele,J_6}t^{\nu\alpha\beta}t_{\mu}{}^{\sigma}{}_\nu(\mathring{\nabla}_\alpha\phi)(\mathring{\nabla}_\beta\phi)(\mathring{\nabla}_\sigma\phi)\,,\nonumber\\
&&-G_{\rm Tele,J_{10}} a^\nu (\mathring{\nabla}_\alpha \phi) (\epsilon^{\mu}{}_{\nu\rho\sigma}t^{\alpha\rho\sigma}+\epsilon^{\alpha}{}_{\nu\rho\sigma}t^{\mu\rho\sigma})\,,\label{Jtele}\\
P_{\phi-\rm Tele}&=&G_{\rm Tele,\phi}\,,\label{PTele}
\end{eqnarray}
to follow the same nomenclature as in standard Horndeski theory and we have also used that $\partial_\mu (eA^\mu) = e\mathring{\nabla}_\mu A^\mu$.\\
The variations of $\sum_{i=2}^{5}\mathcal{L}_i$ with respect to the scalar field are the same as the standard Horndeski equations. According to \cite{Capozziello:2018gms}, one notice that these terms can be written as
\begin{eqnarray}
\delta_{\phi}\Big(e\sum_{i=2}^{5}\mathcal{L}_{i}\Big)=-e\Big[\mathring{\nabla}^\mu\Big(\sum_{i=2}^{5}J^{i}_\mu\Big)-\sum_{i=2}^{5}P_{\phi}^i\Big]\delta \phi\,,\label{phist}
\end{eqnarray}
where $J^{i}_\mu$ and $P^{i}_\phi$ are defined in~\eqref{Pphi}-\eqref{Jis} .

\bibliographystyle{utphys}
\bibliography{references}

\end{document}